\documentclass[iop]{emulateapj}
\usepackage[usenames, dvipsnames]{color}

\slugcomment{Not to appear in Nonlearned J., 45.}

\shorttitle{VVV Survey Microlensing}
\shortauthors{Navarro et al. 2019}

\begin{document}

\title{VVV Survey Microlensing: the Galactic Latitude Dependence}

\author{Mar\'ia Gabriela Navarro \altaffilmark{1,2,3,*} 
Dante Minniti \altaffilmark{1,3,4} 
Joyce Pullen \altaffilmark{3} 
Rodrigo Contreras Ramos \altaffilmark{3,5}}

\affil{Departamento de Ciencias F\'isicas, Facultad de Ciencias Exactas, Universidad Andres Bello, Av. Fernandez Concha 700, Las Condes, Santiago, Chile}
\affil{Dipartimento di Fisica, Universit\`a di Roma La Sapienza, P.le Aldo Moro, 2, I00185 Rome, Italy}
\affil{Millennium Institute of Astrophysics, Av. Vicuna Mackenna 4860, 782-0436, Santiago, Chile}
\affil{Vatican Observatory, V00120 Vatican City State, Italy}
\affil{Instituto de Astrof\'isica, Pontificia Universidad Cat\'olica de Chile, Av. Vicuna Mackenna 4860, 782-0436 Macul, Santiago, Chile}

\begin{abstract}
We search for microlensing events in fields along the Galactic minor axis, ranging from the Galactic center to $-3.7<b< 3.9$ deg., using the VVV survey near-IR photometry. 
The new search is made across VVV tiles $b291$, $b305$, $b319$, $b347$, $b361$ and $b375$, covering a total area of about $11.5$ deg.$^2$.
We find a total of $N=238$ new microlensing events in this new area, $N=74$ of which are classified as bulge red clump (RC) giant sources. 
Combining them with $N=122$ events that we had previously reported in the Galactic center (VVV tile $b333$), allows us to study the latitude distribution of the microlensing events reaching the Galactic plane at $b=0^0$ for the first time.
We find a very strong dependence of the number of microlensing events with Galactic latitude, number that increases rapidly towards the Galactic center by one order of magnitude from $|b|=2$ deg. to $b=0$ deg. with a much steeper gradient than with Galactic longitude. The microlensing event population shows a flattened distribution (axial ratio $b/a \approx 1.5$). The final sample shows a shorter mean timescale distribution than the Galactic plane sample for both, the complete population and RC stars.
\end{abstract}
\keywords{ galaxy: bulge --- galaxy: structure --- gravitational lensing: micro}

\section{Introduction}
\footnotetext{Corresponding author: mariagabriela.navarro$@$uniroma1.it}

Due to the high stellar density, the bulge of the Milky Way is an ideal place to search for microlensing events \citep{Paczynski86}. 
For more than 25 years there have been several optical microlensing experiments that have monitored the Galactic bulge: the Optical Gravitational Lensing Experiment (OGLE; \citealt{Udalski93}), Massive Astrophysical Compact Halo Objects (MACHO; \citealt{Alcock93}), the Microlensing Observations in Astrophysics (MOA; \citealt{Bond01}), the Exp\'erience pour la Recherche d'Objets Sombres (EROS; \citealt{Aubourg93}), the Disk Unseen Objects (DUO; \citealt{Alard95}),  and the Korea Microlensing Telescope Network (KMTNet; \citealt{Kim10}, \citealt{kim17}).

These surveys, however, are incomplete in the Galactic plane because of heavy crowding and extinction. 
\cite{gould95} advocated $K$-band microlensing in the innermost regions of the Galaxy, clearly outlining the advantages of such surveys, especially the ability to observe through the dust found in large quantities in that area.
There are now two new near-IR microlensing searches: one based on the United Kingdom Infrared Telescope Survey (UKIRT, \citealt{Shvartzvald17}) in the Northern hemisphere, and the one based on the VISTA Survey Telescope in the Southern hemisphere, that is carrying out the VVV survey \citep{minniti10,navarro17}.
Both are mapping the inner regions of our Galaxy, that have so far remained hidden from the optical experiments. 

\begin{figure*}[t]
  \includegraphics[width=\textwidth]{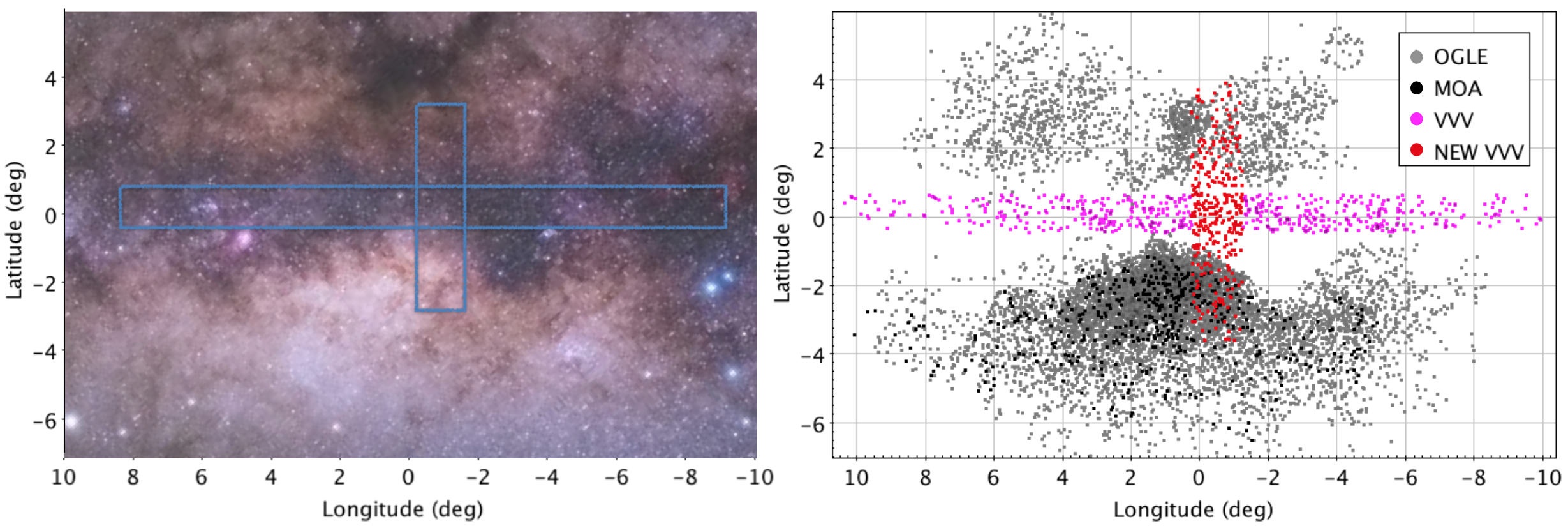}
\caption{Left panel: Areas mapped by \cite{navarro18} along the Galactic plane, and by the present work along the Galactic minor axis, overlaid on the optical image of the Galactic bulge from \cite{mellinger09}.
Right panel: Spatial distribution of the new microlensing events (red squares) above and below the Galactic center. 
The pink squares are the events found in \cite{navarro18}. 
The duplicate events in the overlapping VVV areas have been accounted for. 
The microlensing events from the OGLE EWS and MOA experiments for years 2010 to 2015 are shown for comparison (grey and black squares, respectively). 
The animation showing the final sample of microlensing events along with the light curves is available online. 
In the video the left panel shows the real near-IR image of the area analyzed. 
The flashes reproduce the microlensing events in their specific locations at the relative time when they occurred. 
The duration and intensity of the flashes are proportional to the timescale and amplification, respectively. 
The right panel of the video show the light curve variability with time. 
The colors are linked with the tile where the event were found. 
The colors are shown in the right part of the image, from yellow tones for the most negative latitudes (tile $b291$), to purple tones for the most positive latitudes (tile $b375$)
The realtime duration of the video is 123 seconds representing 5 years of observations.
 \\} 
\label{foto_bulge}
\end{figure*}

The first VVV microlensing events in the Galactic center region were published by \cite{navarro17}, where a clear excess of microlensing events towards the Galactic center was found.
The Galactic longitude dependence of microlensing along the plane was then studied in \cite{navarro18}, where 14 VVV tiles ranging from $l=-10$ to $l=10.44$ deg. were analyse, within a total area covering $20.7$ deg.$^2$ (Figure~\ref{foto_bulge}). 
The final sample was $N = 630$ microlensing events between the years 2010 and 2015, that exhibit a wide and asymmetric distribution with more events towards negative Galactic longitudes, as predicted by the models of \cite{hangould95}, as a consequence of the barred bulge structure.  

\label{sec:sec2}

All models computed for the Galactic bulge predict that the rates of microlensing events --and therefore the optical depth-- increase with decreasing latitude (e.g. \cite{hangould95}, \cite{hangould03}, \cite{wood05}, \cite{ryu08}, \cite{keris09}, \cite{penny13}, \cite{calen14}, \cite{wegg16}, \cite{calenyossi16}, \cite{radek16}, \cite{penny18}).
This reflects the increasing stellar density as one approaches the Galactic plane.
All these models are fine tuned to be in reasonable agreement with the observations from the microlensing optical surveys listed above. 
These surveys, however, do not reach the Galactic plane, extending only to latitudes $|b|>2$ deg., and therefore the predictions for the models at $b=0$ deg. could not be contrasted with observations so far. 

In this paper we complement these studies by dealing with the latitude dependence of microlensing, by analyzing six more VVV tiles.
They cover an area along the Galactic minor axis ($-1.25<l<0.26$ deg.), ranging from $b=-3.7$ to $b=3.9$ deg., making it now possible to bridge the gap from the optical surveys (Figure~\ref{foto_bulge}).
This study of the microlensing latitude dependence is also complementary to the microlensing events studied in the Galactic plane at $b=0$ deg., for $-10<l<10.44$ deg. \citep{navarro18}.

In Section \ref{sec:sec2} the detection procedure and analysis are presented. 
The analysis of the spatial distribution of the new sample and the comparison of our results with other surveys searching in the same area are presented in Section \ref{sec:sec3}. 
The timescale and distance distributions are discussed in Section \ref{sec:sec5}. 
Our final conclusions are presented in Section \ref{sec:sec6}.

\section{Observations and Method}

The PSF photometry has been obtained with DAOPHOT, as described in detail by \citep{contreras17}. 

The microlensing event selection procedure is explained in detail by \cite{navarro17}, \cite{navarro18} and \cite{navarro19}. 
Briefly, the first step consisted in the detection of the best microlensing event candidates using a quality number as an indicator. 
The quality number comprise several features such as a constant baseline, an increase and decrease in brightness in a symmetrical way, and the $\chi^2$ of the fit. 
In this step we also rejected from the final sample the light curves with less than $N=20$ data points and the light curves with less than $N=4$ data points during the event. 
Approximately $20,000$ light curves meet these initial criteria. 
The second step was to visually inspect the events with the best quality number to obtain the final sample. 
This procedure leave out of the sample the binary events and events with strong parallax effect. 
The final sample for the six new tiles analyzed consists in $N=238$ microlensing events (Figure~\ref{foto_bulge}). 
We found that $N= 70$ events were previously detected by OGLE in the overlapped region.

\begin{figure*}[t]
\epsscale{1.2}
\plotone{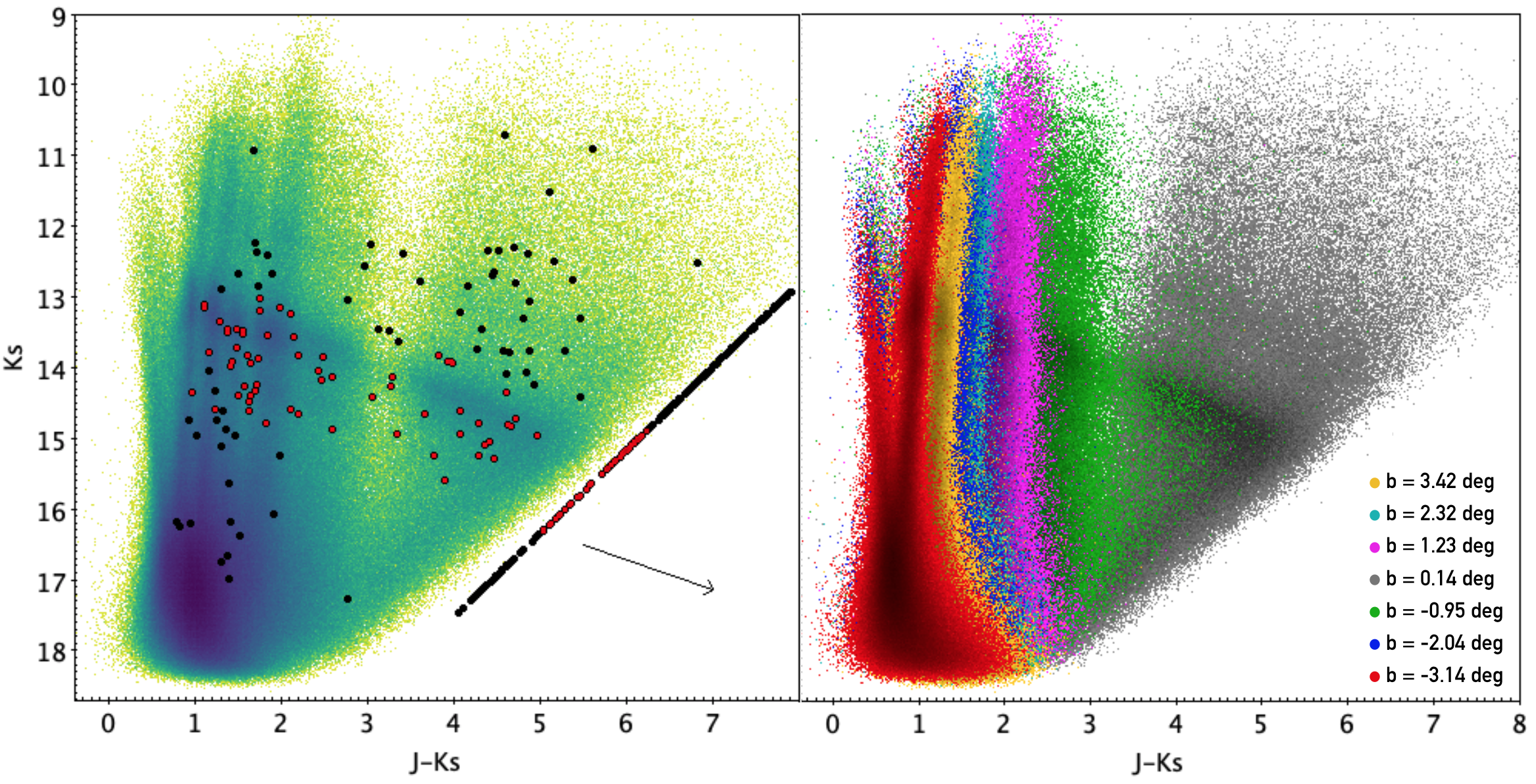}
\caption{Near-IR $K_s$ vs $J-K_s$ color-magnitude diagrams for the 7 VVV tiles (from $b327$ to $b340$). 
Left panel: A logarithmic color-coded Hess diagram representation of the $\sim 40$ million individual sources. 
The black circles indicate the sources of the sample of microlensing events, and the red circles show the selected RC sources. 
The line of points to the right of the CMD is due to the events that are only detected in the $K_s$-band and have no color information, because the sources are too red (i.e. fainter than the detection limits in the $J$ and $K_s$-bands).
The arrow show the extinction law proposed by \cite{alonso18} specially computed for the VVV Survey.
Right panel: CMDs of the different tiles analysed plotted in different colors along with their central latitudes.
The central longitude of all the tiles is $l \sim -0.5$ deg. 
The dimensions of the tiles are $1.5$ deg. and $1.2$ deg. for longitude and latitude respectively. \\}
\label{cmd_old}
\end{figure*}

It is more convenient to restrict the sample to RC sources, as they are usually less affected by incompleteness and describe better the microlensing properties in the Milky Way bulge (e.g. \cite{gould95}, \cite{pop01}, \cite{pop05}, \cite{afonso03}).
Figure~\ref{cmd_old} shows the near-IR $K_s$ vs $J-K_s$ color-magnitude diagram (CMD) for the 7 VVV tiles ($b291$, $b305$, $b319$, $b333$, $b347$, $b361$ and $b375$) covering a range of latitudes along the Galactic minor axis. 
Comparing this CMD with that of the Galactic plane (Figure 8 of \cite{navarro18}), the most striking difference is that reddening is much reduced in these fields above and below the Galactic plane. 
In order to highlight the extreme and irregular reddening of the plane we show the CMD of each tile plotted in different colors (right panel of Figure~\ref{cmd_old}). 
The CMD of the innermost tile $b333$ is shown in grey and is the most affected by extinction as expected.
 The red clump of every CMD is clear and follow the extinction law proposed by \cite{alonso18}.

Left panel of Figure~\ref{cmd_old} shows the position of the microlensing sources, outlining the selected RC events. 
For sources fainter than the detection limit in $J$ band, the color was estimated using the color limit in the CMD. 
When the $K_s$ magnitude from the photometry was not well constrained, we use the baseline magnitude of the $K_s$ light curve.
This is the reason why a line of events are formed in the red side of the CMD.
The color cut for the red clump selection depend on the tile, due to the different extinction present in each of them.

We use $J-K_s$ in this case but using the $H-K_s$ CMD the results remain the same.
Table~\ref{tbl-1} lists the Galactic coordinates of each tile along the Galactic minor axis, along with the number of total light-curves analysed, total RC sources analyzed, number of events found, and the corresponding numbers of RC events. All of them are presented as raw values and corrected by completeness.

\section{Spatial Distribution along the Galactic Minor axis and comparison with other surveys}
\label{sec:sec3}

\begin{figure}[t]
\epsscale{1.2}
\plotone{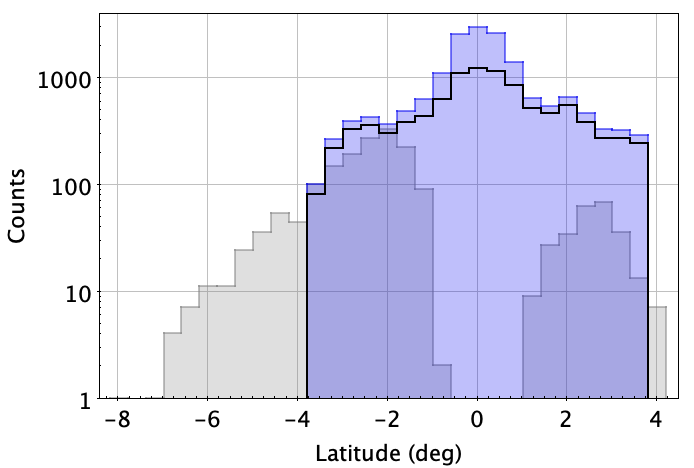}
\caption{Observed distribution of the total number of microlensing events. 
OGLE distribution of events found between 2010 and 2015 is shown un grey.  
The black line shows the raw VVV distribution arbitrarily normalized to the Southern OGLE results in the overlap regions, that have more events. 
The blue histogram is the distribution corrected by completeness. \\}
\label{comparison}
\end{figure}

\begin{figure*}[t]
  \includegraphics[width=\textwidth]{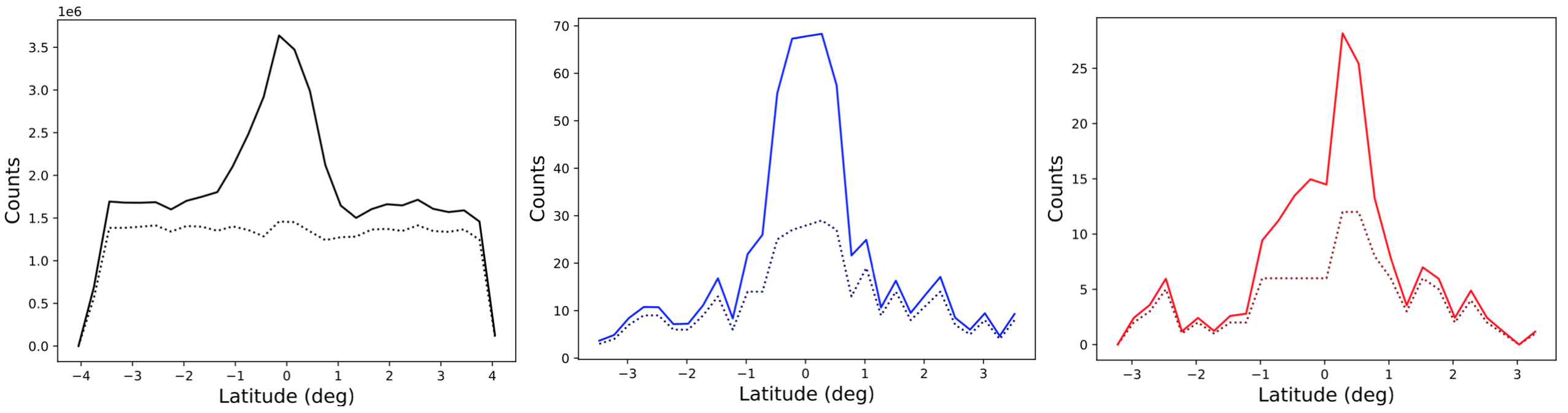}
\caption{
Left: Distribution of the total number of stars ($N_{*s}$) as a function of latitude.
Center: Distribution of the total number of microlensing events ($N_{\mu l}$) as function of Galactic latitude.
Right: Distribution of the red clump sample  ($N_{\mu l, RC}$) along the Galactic latitude. 
In all the cases, the dotted line is the raw distribution and the filled line is the completeness corrected distribution ($N_{*}^{corr} $, $N_{\mu l}^{corr}$, $N_{\mu l, RC}^{corr}$, respectively). \\
 \label{nevents}}
\end{figure*}

The spatial distribution of the complete sample is shown in Figure~\ref{foto_bulge}, along with the distribution of events discovered by the OGLE Early Warning System (EWS\footnote{http://ogle.astrouw.edu.pl}) and MOA optical surveys between years 2010 and 2015 \citep{udalski15, sumi13}. This figure underscores the importance of the VVV survey that completes the microlensing census at low latitudes, where the optical surveys are blind because of the high extinction.

It is important to check the number of events compared with OGLE and MOA, in order to obtain an external completeness estimate.
Although these different microlensing experiments are not directly comparable, as they all have very  different depths, wavelength coverage, and time sampling.
    
The areal overlap between VVV and OGLE is non uniform, covering approximately $4$ deg.$^2$ above and below the plane ($\sim 8.25$deg.$^2$ in total). 
In this common area OGLE detected $N=1567$ events using the same time baseline (2010 to 2015), while we found $N=179$ events in total, $N=109$ of them are new discoveries, generally located in regions of high extinction where OGLE has fewer observations.
This order of magnitude difference in the number of events is not surprising, considering that OGLE observed by more than an order of magnitude more nights than the VVV survey, with a total mean sampling of about $N=85$ nights. Additionally, the VVV bulge observations are random, with typical spacing of a few days (1-3 days), but this is heavily field dependent, and of course there are gaps due to weather and the Moon (some fields cannot be observed once a month because the Moon moves along the ecliptic that goes through the bulge), and seasonal gaps \citep{hempel14}.
Therefore, on average both microlensing experiments reported $\sim 2$ events per night. 
But this still means that the VVV survey is only detecting $\sim 10\%$ of the events. 

This is important to take into account when designing the microlensing experiment for WFIRST (\cite{Green12}, \cite{Spergel15}): the potential number of events to be discovered by WFIRST may largely exceed previous expectations if it observes the Galactic center fields.
For example, using our measured (raw) latitude dependence distribution, normalised to fit the Southern OGLE region that have more events
(black histogram of Figure~\ref{comparison}), we find that WFIRST would detect more than $N=7,000$ events in the Galactic center fields i.e. $|b|\sim 2$ deg. (many more considering that it would be even more efficient than OGLE). Off course this depends on the observational setup, for which it would be very advantageous to extend the WFIRST's filter from $1.0-2.0 \mu m$ to $1.0-2.4 \mu m$ \citep{penny18} or to include a $K_s$-band filter \citep{Stauffer18}.

Conversely, in the overlap area with MOA that spans about $2$ deg.$^2$ they detected $N=54$ events, while we find $N=56$ events. 
They are detecting on average about two events per night.

To make a more accurate analysis of the spatial distribution and its dependence with latitude it is important to consider the completeness corrected distributions. 
The correction for sampling efficiency does not alter the observed distributions, as the relative number of observations are relatively similar for all VVV survey tiles (e.g. Figure 23 of \cite{navarro19}). 
However, the correction for photometric efficiency is critical, as we move closer to the Galactic center. 
We apply the completeness corrections following \cite{navarro19}, that are based on the extensive artificial star simulations of \cite{hempel15} and \cite{valenti16} using the PSF photometry of \cite{alonso18}. 
Considering the completeness map for RC stars for stars down to $K_{s0} = 14$ mag, for example, these corrections increase from $\sim 10\%$ for tiles $b375$ and $b291$ at $|b|=3$ deg., to $\sim 50\%$ at the Galactic center tile $b333$ at $|b|=0$ deg. 
The VVV detection limit is field dependent, ranging from $K_s \sim 18$ mag in most fields, to $K_s \sim 16.5$ mag in the Galactic center tile \citep{saito12}.

The completeness corrected distribution of the sample is shown in Figure~\ref{comparison} (blue histogram). 
Using the completeness distribution we expect to find more than $N=13,000$ events at $|b|\sim 2$ deg.

Figure~\ref{nevents} shows the Galactic latitude distribution of the total number of stars (black), microlensing events (blue) and the number of RC events (red).
The observed distributions are shown as dashed lines, while the completeness corrected distributions are plotted as filled lines. 

The distribution of the 35 million stars analyzed (left panel of Figure~\ref{nevents}) show an increase in the number density of stars that is more peaked when the completeness correction is applied. 
From Figure~\ref{nevents} we see that the number of microlensing events clearly increases with decreasing latitude, reaching a maximum in the Galactic plane at $b=0$ deg.
There are $N=181$ total observed events within $-1<b<1$ deg. that is the $50 \%$ of the sample, while on average there are $N=89$ observed events  with $1<|b|<3$ deg., a factor of $2\times$ less. 
The respective numbers of events corrected for completeness are $N \sim 390$ within $-1<b<1$ deg., and $ N \sim 100$ within $1<|b|<3$ deg., i.e. a factor of within $3.9\times$ less.
 
The effect is even more dramatic if we consider only the RC events, which should be a more complete and unbiased sample in comparison. 
There are $N=62$ total observed events within $-1<b<1$ deg., while on average there are $N=25$ observed events with $1<|b|<3$ deg., a factor of $2.5\times$ less.
The respective numbers of events corrected for completeness are $N \sim 130$ within $-1<b<1$ deg., and $N \sim 30$ within $1<|b|<3$ deg., i.e. a factor of within $4.4\times$ less.

\begin{figure}
\epsscale{1.2}
\plotone{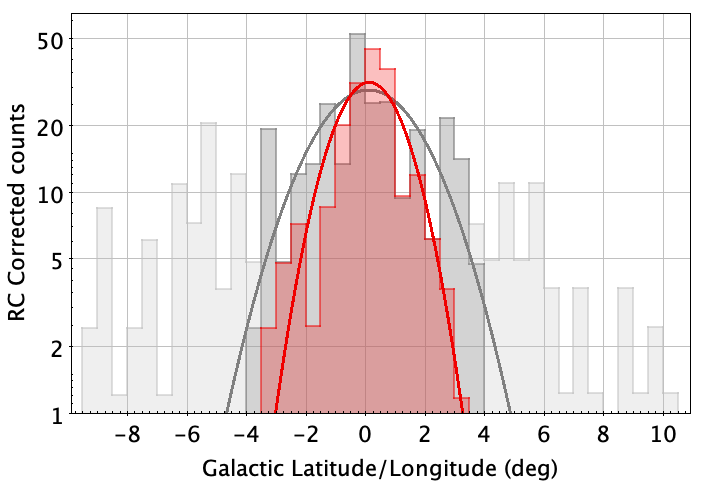}
\caption{Corrected RC microlensing event distribution along Galactic latitude (red histogram) compared with the corrected distribution of RC events along Galactic longitude (grey histogram) from \cite{navarro18}. 
The lines are the Gaussian fit of each distribution. The fit was done within $-3.7<l< 3.9$ deg. and $-3.7<b< 3.9$ deg. 
The $FWHM = 2.82$ deg. and $FWHM = 4.32$ deg. along latitude and longitude samples, respectively. \\
\label{compare}}
\end{figure}

This trend is even clearer in Figure~\ref{compare} where the corrected distribution of the red clump sample is compared with the microlensing events with RC sources found in \cite{navarro18} for the longitude dependence studies. 
In both cases the number of microlensing events increase towards the Galactic center, but the distribution is wider for the longitude dependence and more pronounced for the latitude dependence. 
The distribution of the Galactic minor axis sample is fairly symmetric, and more concentrated,  while the distribution of the Galactic plane sample is more asymmetric and wider.
These distributions have $FWHM = 2.82$ deg. and $FWHM=9.09$ deg. along latitude and longitude, respectively. 
In order to be consistent, we limited the sample along the plane within the same area analysed for the latitude dependence($-3.7<l< 3.9$ deg.). 
In this case the distribution has a $FWHM = 4.32$ deg. reflecting that the microlensing events have a flattened distribution (axial ratio $b/a \approx 1.5$) in the innermost regions of the Milky Way, explored here for the first time.

To summarize, we find that the completeness corrected distribution is more peaked, with the number of events strongly increasing when we move from 2 degrees latitude to zero latitude by a factor of an order of magnitude. 
As far as we have explored the literature, none of the existing models previously predicted such a large effect.

This has important implications for the microlensing plans for the Wide Field Infrared Space Telescope (WFIRST)  mission \citep{Spergel15}, because most of the selected fields for the WFIRST microlensing campaign are located at $b=-2$ deg. \citep{penny18}, thereby missing an important increment in the number of potentially detectable events in comparison with a campaign done at $b=0$ deg. Additionally, as already pointed out, in order to take better advantage of a microlensing survey in the Galactic plane, an extension of the WFIRST's filter from $1.0-2.0 \mu m$ to $1.0-2.4 \mu m$ \citep{penny18} or even better, a $K_s$-band filter its necessary onboard the WFIRST \citep{Stauffer18}.

\section{Timescale and Distance Distributions}
\label{sec:sec5}

\begin{figure}
\epsscale{1.2}
\plotone{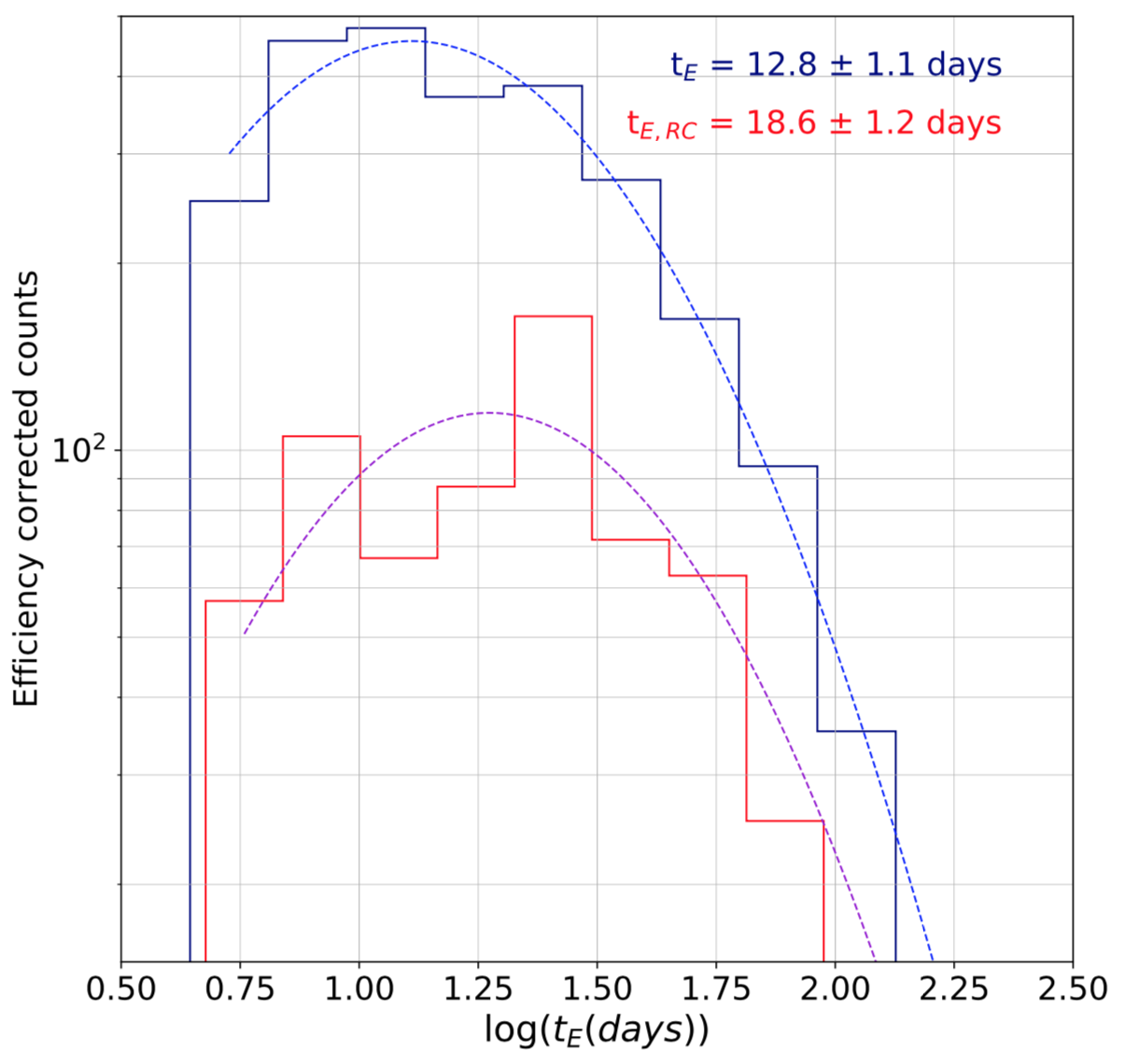}
\caption{Efficiency corrected timescale distribution for the microlensing sample along the minor axis (blue histogram) and the RC events (red histogram). 
The mean timescale for the complete sample is $t_E = 12.8 \pm 1.1$ days, while for the RC sample is $t_E = 18.6 \pm 1.2$ days.  
This distribution exclude the innermost tile $b333$. \\
\label{tscaleb}}
\end{figure}

The efficiency corrected timescale distribution obtained for the present sample is shown in Figure~\ref{tscaleb}, for both the whole sample and the selected RC sample. 
The mean timescale for the present sample of events along the Galactic minor axis (excluding the Galactic center i.e. tile $b333$) is $t_E = 12.8 \pm 1.1$ days, while for the RC sample is $t_E = 18.6 \pm 1.2$ days.  
The completeness correction was computed using fake microlensing light curves and is explained in detail in \cite{navarro19}. 

As another check for our sample, we can compare the timescale distribution for the events in the new tiles with our previous results published by \cite{navarro18} along the Galactic plane at $b=0$ deg. 
The events along that Galactic plane have a mean timescale of $17.4 \pm 1.0$ days for the whole sample of $N=630$ events, and $20.7 \pm 1.0$ days for the selected RC sample of $N=291$ events, respectively \citep{navarro18}.

\begin{figure}
\epsscale{1.2}
\plotone{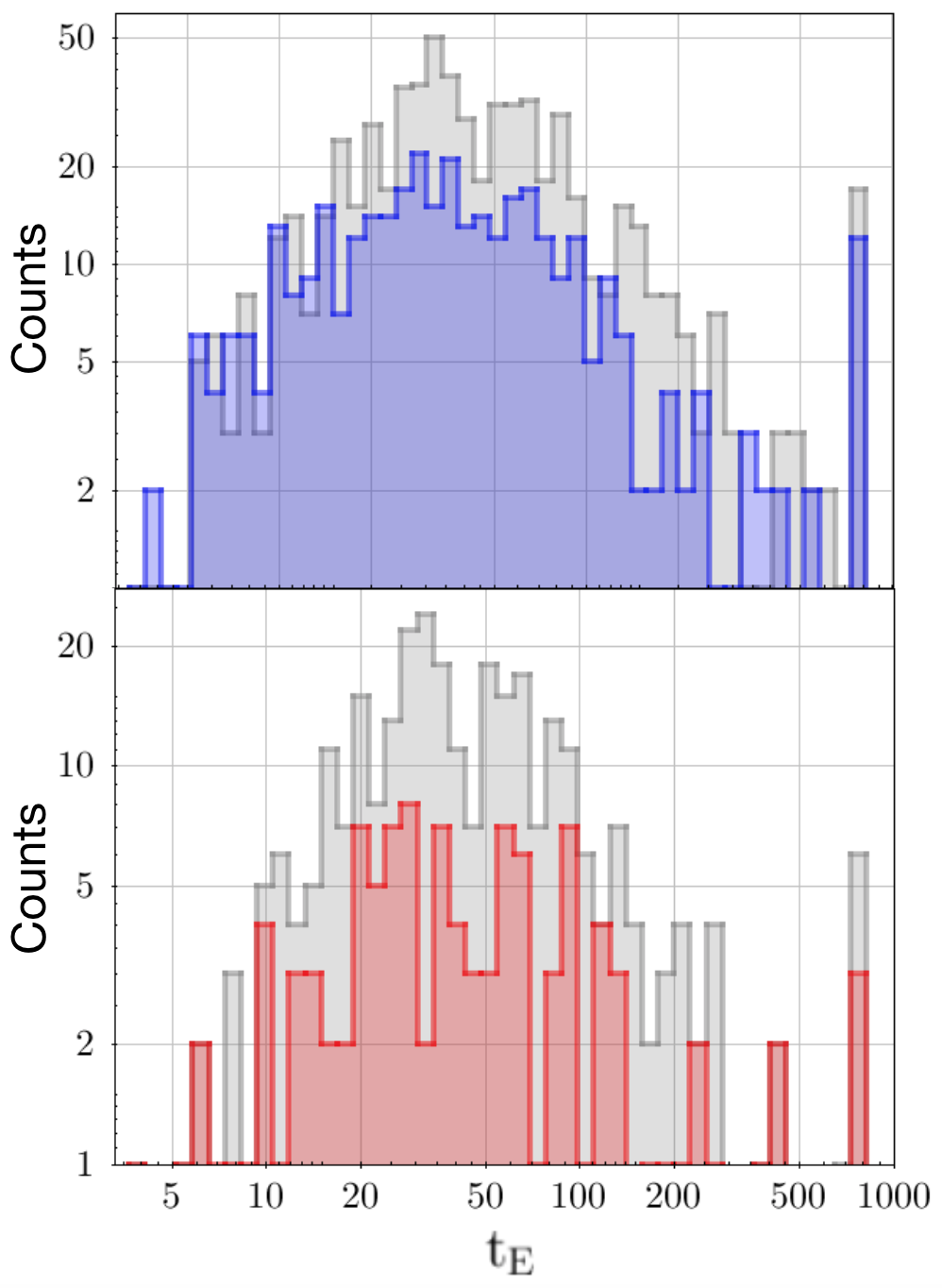}
\caption{Comparison between the raw timescale distribution for the samples along the Galactic minor axis (blue and red histogram), and major axis (grey histogram, from \cite{navarro18}). \\
\label{tscalea}}
\end{figure}

Figure~\ref{tscalea} shows the comparison between the raw timescale distribution for the complete sample (blue) and RC sample (red) along the Galactic minor axis and the sample along the Galactic plane from \cite{navarro18} (grey). 
In both cases is clear the shift in the mean timescale distribution towards shorter timescales. 

These observed differences in the mean timescales between the samples along the Galactic minor axis and the Galactic plane are significant at the $4-5 \sigma$ level. 
Subtle differences may be expected because the distance distribution of the sources (and therefore the lenses) should be different in these two samples, as the line of sight departs from the Galactic plane. 

In both cases (raw and efficiency corrected timescale distributions) the shape of the timescale distribution along the Galactic minor axis (excluding the Galactic center tile i.e. $b333$) is in good agreement with our previous study of the timescale distribution along the Galactic plane.
However, the minor axis sample events have a shorter timescale in the mean than the Galactic plane sample events. This difference is more pronounced for the complete sample than for the RC subsample.

\begin{figure}
\epsscale{1.2}
\plotone{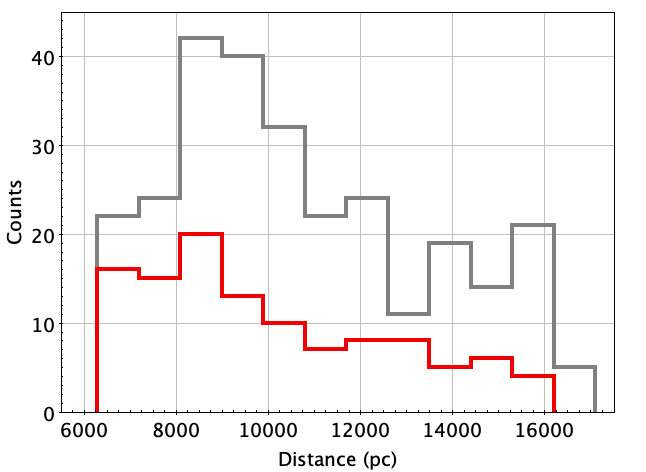}
\caption{Observed RC microlensing event distance distribution along Galactic latitude (red histogram) compared with the distance distribution of the total number of RC events along Galactic longitude (grey histogram) from \cite{navarro18}.  \\
\label{dist}}
\end{figure}

Indeed, Figure~\ref{dist} shows the comparison of the distance distributions for the RC sources in these two samples.
The RC distances are computed using the Gaia RC mean magnitudes and colors from \cite{ruiz18}, following the method of \cite{navarro19}.
Note that we adopt the mean intrinsic magnitude of the RC to be $K_{s0}= -1.61\pm 0.01$ \citep{ruiz18}, but adopting a different calibration such as 
$K_{s0}= -1.68\pm 0.03$ from \cite{alves02} or even computing the individual extinctions for all the RC sources using the last version of the BEAM calculator \citep{gonzalez12} does not change our conclusions. 

Figure~\ref{dist} shows that the Galactic minor axis RC sources (red histogram) are mostly located at the distance of the Galactic bulge ($6<D<12$ kpc), as expected.
On the other hand, the Galactic plane RC sources (grey histogram) are more asymmetric and widely distributed. 
Their distance distribution peals at the distance of the Galactic center, but a sizeable fraction of the population ($20-30\%$) is clearly located in the background, beyond $D\sim 12$ kpc, including some located at the Galactic antipodes ($D\sim 16$ kpc).

Figure~\ref{dist} clearly shows that some of the sources of the low latitude microlensing events may be very distant, demonstrating the usefulness of the VVV survey for probing the populations much beyond the Galactic center. 
However, larger numbers of microlensing events and detailed modeling are needed in order to explore in detail whether the observed difference in distances can explain the difference in timescales between the two samples.

\section{Conclusions}
\label{sec:sec6}
We have detected $N=238$ microlensing events within an area of $11.5$ deg.$^2$ along the Galactic minor axis at $l=-0.5$ deg. using the near-IR VVV Survey photometry. 

In \cite{navarro18} we found that there is an excess of microlensing events at negative longitudes, and that the observed longitude dependence is very shallow, as expected from the model predictions. 
The latitude dependence however, should be stronger, as also predicted by the models and observed in the higher latitude fields by the optical microlensing surveys.

We find that the Galactic latitude dependence is much stronger than the longitude dependence, which is significantly shallower. 
The observed number of events drops from 
$N=122$ at $b333$ to $N\sim68$ at $b332$ and $b334$ at $1.5$ deg. in longitude (decrease by $41\%$), 
(a crude interpolation yields at $1.1$ deg. in latitude a decrease by $31\%$)
while they drop from 
$N=122$ at $b333$ to $N=59$ and $N=56$ at $b319$ and $b347$ respectively, at $1.1$ deg. in latitude (decrease by $47\%$).

The completeness corrected trends highlight the difference more strongly. 
The differential extinction was taken into account on a field by field basis, if there is any residual effect it would be in the sense of making the real gradient even more pronounced. 
In other words, because extinction is more severe as one approaches the Galactic plane, the number of events in these regions should be larger than observed. 

The $FWHM$ of the distribution is 1.5$\times$ larger along Galactic longitude than along latitude, indicating a highly flattened microlensing event rate distribution in the innermost regions of the Milky Way.

In summary, we find that there is an order of magnitude more microlensing events at $b=0$ deg. than at $|b|=2$ deg. 
This is significantly higher than all the expectations from all the previous models (that were fine tuned to explain the observations from the optical surveys at higher latitudes).
The higher event rate observed in the Galactic plane is probably due to a combination of two effects: the higher source density, and the higher microlensing optical depth.

We also find that the mean timescales of the Galactic plane sample from \cite{navarro18} is longer than the present sample along the Galactic minor axis, both considering the total samples or the selected RC giant star events. 
These differences may be due to the different distance distribution of the sources of the microlensing events (and therefore also of the lenses).

We find that the present sample of events along the Galactic minor axis contains fewer distant RC giants located beyond the bulge than the Galactic plane sample of \cite{navarro18}, where the distant RC giants are much more numerous.
We argue that larger samples of microlensing events in the direction of the bulge, and detailed modeling are needed to study these trends.

This is the first time a latitude analysis of the microlensing event population is done reaching the Galactic plane at $b=0$ deg. 
This study enables more complete future modeling of the microlensing rates and optical depths throughout the Galactic bulge. 
We also strongly advocate for a WFIRST microlensing campaign along the Galactic plane, where the event rate is clearly maximal.

\acknowledgments
We gratefully acknowledge the use of data from the ESO Public Survey program IDs 179.B-2002 and 198.B-2004 taken with the VISTA telescope and data products from the Cambridge Astronomical Survey Unit. 
Support for the authors is provided by the BASAL Center for Astrophysics and Associated Technologies (CATA) through grant AFB 170002, by the Programa Iniciativa Cientifica Milenio grant IC120009, awarded to the Millennium Institute of Astrophysics (MAS), and by Proyecto FONDECYT No. 1170121.



\begin{deluxetable}{rrrrrrrrrrrrrr}
\tablecaption{
Number of VVV Survey Microlensing Events Along the Galactic Minor Axis. 
The tiles are presented along with the equatorial and Galactic coordinates, the number of sources and red clump sources analyzed and the number of events for the whole sample and the red clump subsample, both raw and corrected by completeness. The subscript ``F'' indicates that the real number is obtained by multiplying the value by a factor of $\times 10^5$.
\label{tbl-1}}
\tablehead{
\colhead{Tile} & \colhead{RA(J2000)}  & \colhead{DEC(J2000)} & \colhead{l(deg.)}& \colhead{b(deg.)} & \colhead{$N_{*,F}$} & \colhead{$N_{*,F}^{corr} $}  & \colhead{$N_{*RC,F}$}   & \colhead{$N_{*RC,F}^{corr}$}  & \colhead{$N_{\mu l}$} & \colhead{$N_{\mu l}^{corr}$}  & \colhead{$N_{\mu l, RC}$} & \colhead{$N_{\mu l, RC}^{corr}$}
}
\startdata
$b291$ &17 56 53.736 &-30 58 01.200   & 359.50054  &-3.13672   & 53.7 & 65.2 & 3.8 & 4.7 & 22 & 26 & 5 & 6 \\
$b305$ &17 52 29.424 &-30 25 22.080   & 359.49334  &-2.04452   & 49.2 & 59.2 & 6.3 & 7.6 & 34 & 40 & 10 & 12 \\
$b319$ &17 48 08.616 &-29 51 58.320   & 359.48996  &-0.95233   & 48.9 & 77.8 & 10.8 & 17.3 & 59 & 93 & 18 & 28 \\
$b333$ &17 43 51.192 &-29 17 52.800   & 359.48985  & 0.13988  & 60.6 & 144.5 & 13.7 & 32.7 & 122 & 290 & 38 & 90 \\
$b347$ &17 39 37.152 &-28 43 06.240   & 359.49322  & 1.23206  & 53.8 & 64.4 & 9.4 & 11.2 & 56 & 66 & 25 & 29 \\
$b361$ &17 35 26.472 &-28 07 39.360   & 359.50024  & 2.32421  & 58.4 & 71.3 & 5.8 & 7.0 & 42 & 51 & 14 & 17 \\
$b375$ &17 31 19.032 &-27 31 34.680   & 359.51046  & 3.41635  & 58.3 & 67.8 & 3.4 & 3.9 & 25 & 29 & 2 & 2 \\
\hline
\enddata
\end{deluxetable}

\begin{longtable*}{ l l l l l l l l l l l l}
\tablecaption{VVV Survey first quality microlensing events with their respective positions in Galactic coordinates, baseline $K_s$ magnitude, Color and the parameters obtained using the standard microlensing model including the blending ($f_{bl}$). The events with DET were detected by OGLE.
} \label{tab:table1} \\
\tablehead{
\colhead{Tile} & \colhead{ID} & \colhead{l} & \colhead{b} & \colhead{$K_s$} & \colhead{$J-K$} & \colhead{Amp} & \colhead{$u_{0}$} & \colhead{$t_{0}$} & \colhead{$t_E$} & \colhead{$f_{bl}$} & \colhead{Comment}   \\
 & &  {\tiny (deg.)} &  {\tiny (deg.)} & {\tiny (mag)} & {\tiny (mag)}  & &  & {\tiny (MJD)} & {\tiny (days)}  &  & \\ }

291 & 14359 & -0.83423 & -3.56950 & 14.34 & 6.71 & 1.42 & 0.36 $\pm$ 0.70 & 57224.73 $\pm$ 16.88 & 71.50 $\pm$ 85.47 & 0.92 $\pm$ 2.71 &   \\
291 & 18483 & -0.76825 & -3.57744 & 14.23 & 6.80 & 0.38 & 0.53 $\pm$ 0.60 & 55811.60 $\pm$ 1.81 & 37.70 $\pm$ 26.38 & 0.39 $\pm$ 0.67 & DET. \\
291 & 60869 & -1.16721 & -3.28885 & 17.46 & 4.06 & 12.64 & 0.06 $\pm$ 0.05 & 56794.68 $\pm$ 0.35 & 115.33 $\pm$ 57.01 & 1.00 $\pm$ 0.70 & DET. \\
291 & 8466 & -0.82319 & -2.63851 & 14.13 & 6.89 & 0.44 & 0.86 $\pm$ 1.06 & 56908.31 $\pm$ 4.48 & 69.65 $\pm$ 55.27 & 1.00 $\pm$ 2.37 & DET. \\
291 & 95627 & -0.85515 & -3.34676 & 16.20 & 5.13 & 5.32 & 0.00 $\pm$ 2.00 & 56738.29 $\pm$ 0.17 & 750.00 $\pm$ 1172.43 & 0.02 $\pm$ 0.03 &   \\
291 & 58006 & -0.44430 & -2.72898 & 13.09 & 7.77 & 0.13 & 0.00 $\pm$ 2.00 & 56869.67 $\pm$ 2.48 & 130.01 $\pm$ 90.84 & 0.02 $\pm$ 0.02 & DET. \\
291 & 59043 & -0.25578 & -2.73065 & 13.30 & 7.60 & 0.29 & 0.83 $\pm$ 2.00 & 56776.70 $\pm$ 2.71 & 16.77 $\pm$ 35.20 & 1.00 $\pm$ 7.04 & DET. \\
291 & 71521 & -0.25151 & -2.75399 & 13.00 & 7.85 & 3.23 & 0.16 $\pm$ 0.03 & 55803.04 $\pm$ 0.28 & 116.91 $\pm$ 14.52 & 0.90 $\pm$ 0.18 & DET. \\
291 & 31045 & -1.00208 & -3.60150 & 12.66 & 8.14 & 0.22 & 1.20 $\pm$ 1.72 & 57094.94 $\pm$ 7.52 & 149.73 $\pm$ 143.03 & 1.00 $\pm$ 3.27 & DET. \\
291 & 33760 & -0.45070 & -3.23385 & 17.39 & 4.11 & 1.78 & 0.27 $\pm$ 0.71 & 56524.65 $\pm$ 4.78 & 44.40 $\pm$ 84.13 & 0.60 $\pm$ 2.09 & DET. \\
291 & 48672 & 0.00268 & -2.72106 & 15.41 & 5.80 & 0.77 & 0.00 $\pm$ 2.00 & 56819.98 $\pm$ 4.27 & 750.00 $\pm$ 4258.06 & 0.01 $\pm$ 0.04 &   \\
291 & 79253 & -0.07317 & -3.32179 & 13.15 & 7.72 & 1.98 & 0.32 $\pm$ 0.02 & 56123.10 $\pm$ 0.04 & 13.95 $\pm$ 0.63 & 0.90 $\pm$ 0.08 & DET. \\
291 & 9436 & 0.07983 & -2.64175 & 16.74 & 4.67 & 4.47 & 0.08 $\pm$ 0.09 & 56113.04 $\pm$ 0.80 & 13.89 $\pm$ 9.56 & 0.41 $\pm$ 0.49 & DET. \\
291 & 31223 & -0.13540 & -3.59885 & 17.08 & 4.38 & 1.46 & 0.00 $\pm$ 2.00 & 56811.82 $\pm$ 2.08 & 31.62 $\pm$ 27.70 & 1.00 $\pm$ 1.72 &   \\
291 & 27754 & 0.19082 & -3.04078 & 15.13 & 6.04 & 1.09 & 0.41 $\pm$ 0.21 & 56013.86 $\pm$ 4.04 & 125.65 $\pm$ 37.42 & 0.67 $\pm$ 0.48 & DET. \\
291 & 36659 & 0.20795 & -3.05757 & 15.94 & 5.35 & 2.37 & 0.28 $\pm$ 0.24 & 56557.06 $\pm$ 0.48 & 21.66 $\pm$ 11.19 & 1.00 $\pm$ 1.03 & DET. \\
291 & 40922 & 0.14566 & -3.06548 & 17.19 & 4.29 & 2.52 & 0.16 $\pm$ 0.25 & 56153.07 $\pm$ 1.08 & 33.18 $\pm$ 28.89 & 0.61 $\pm$ 0.91 & DET. \\
291 & 41165 & -0.73302 & -3.06388 & 13.74 & 7.22 & 0.60 & 0.65 $\pm$ 0.32 & 56448.53 $\pm$ 1.28 & 107.52 $\pm$ 30.31 & 1.00 $\pm$ 0.78 & DET. \\
291 & 42518 & -1.05530 & -2.88447 & 13.24 & 7.64 & 0.22 & 0.00 $\pm$ 2.00 & 56809.20 $\pm$ 9.24 & 12.15 $\pm$ 53.86 & 1.00 $\pm$ 13.27 &   \\
291 & 43816 & -0.47510 & -2.88488 & 11.97 & 8.73 & 2.38 & 0.00 $\pm$ 2.00 & 56402.71 $\pm$ 4.54 & 351.68 $\pm$ 97.48 & 0.15 $\pm$ 0.06 & DET. \\
291 & 15478 & -0.08828 & -2.83466 & 15.26 & 5.93 & 1.01 & 0.46 $\pm$ 0.48 & 56149.96 $\pm$ 0.59 & 25.12 $\pm$ 15.45 & 0.86 $\pm$ 1.26 & DET. \\
291 & 87991 & -0.02270 & -2.97132 & 13.32 & 7.58 & 0.74 & 0.66 $\pm$ 0.20 & 56153.43 $\pm$ 0.28 & 33.58 $\pm$ 6.57 & 1.00 $\pm$ 0.51 & DET. \\
305 & 51721 & -1.13552 & -1.99593 & 13.59 & 7.34 & 0.34 & 0.56 $\pm$ 2.00 & 56149.91 $\pm$ 0.42 & 6.60 $\pm$ 13.10 & 1.00 $\pm$ 6.45 & DET. \\
305 & 58739 & -0.69976 & -2.55682 & 15.19 & 5.99 & 1.62 & 0.07 $\pm$ 0.05 & 56840.22 $\pm$ 0.09 & 18.99 $\pm$ 11.10 & 0.13 $\pm$ 0.09 & DET. \\
305 & 102736 & -0.82319 & -2.63851 & 14.11 & 6.90 & 0.43 & 0.86 $\pm$ 0.87 & 56910.29 $\pm$ 4.67 & 78.95 $\pm$ 49.93 & 1.00 $\pm$ 1.93 &   \\
305 & 5363 & -0.86367 & -1.54147 & 13.59 & 7.35 & 0.32 & 1.02 $\pm$ 2.00 & 55811.96 $\pm$ 1.92 & 32.05 $\pm$ 71.55 & 1.00 $\pm$ 7.03 &   \\
305 & 28573 & -0.82439 & -2.13497 & 13.83 & 7.15 & 0.26 & 0.11 $\pm$ 2.00 & 56880.04 $\pm$ 5.65 & 18.52 $\pm$ 106.53 & 1.00 $\pm$ 21.02 & DET. \\
305 & 44942 & -0.34826 & -1.61946 & 14.06 & 6.95 & 0.32 & 0.27 $\pm$ 1.98 & 56534.28 $\pm$ 2.31 & 47.94 $\pm$ 217.16 & 0.15 $\pm$ 1.28 & DET. \\
305 & 89982 & -1.23501 & -1.89040 & 13.59 & 7.35 & 3.95 & 0.20 $\pm$ 0.03 & 56117.73 $\pm$ 0.10 & 25.88 $\pm$ 2.40 & 1.00 $\pm$ 0.16 &   \\
305 & 40818 & -0.73734 & -1.79001 & 16.70 & 4.70 & 2.53 & 0.19 $\pm$ 0.33 & 56369.75 $\pm$ 0.26 & 21.58 $\pm$ 30.71 & 0.62 $\pm$ 1.29 & DET. \\
305 & 72856 & -0.81026 & -1.85066 & 16.87 & 4.56 & 2.01 & 0.23 $\pm$ 0.70 & 56863.31 $\pm$ 6.36 & 41.16 $\pm$ 70.79 & 1.00 $\pm$ 3.48 & DET. \\
305 & 27319 & -0.87610 & -2.31628 & 12.63 & 8.16 & 0.40 & 0.86 $\pm$ 2.00 & 55848.44 $\pm$ 0.84 & 14.60 $\pm$ 64.20 & 1.00 $\pm$ 14.33 &   \\
305 & 64527 & -1.01654 & -1.83685 & 13.05 & 7.81 & 0.15 & 0.00 $\pm$ 2.00 & 55927.89 $\pm$ 2.98 & 106.95 $\pm$ 74.85 & 0.16 $\pm$ 0.30 &   \\
305 & 40638 & -0.91480 & -2.34845 & 13.90 & 7.08 & 2.05 & 0.29 $\pm$ 0.04 & 56096.17 $\pm$ 0.34 & 61.13 $\pm$ 5.51 & 0.82 $\pm$ 0.14 & DET. \\
305 & 67899 & -0.61263 & -1.83984 & 14.46 & 6.61 & 5.04 & 0.11 $\pm$ 0.01 & 55808.83 $\pm$ 0.10 & 87.81 $\pm$ 6.58 & 1.00 $\pm$ 0.11 & DET. \\
305 & 66801 & -0.56543 & -2.38898 & 15.33 & 5.87 & 2.70 & 0.27 $\pm$ 0.38 & 55788.67 $\pm$ 2.30 & 31.04 $\pm$ 23.15 & 1.00 $\pm$ 1.66 & DET. \\
305 & 75109 & -0.67287 & -2.40470 & 13.72 & 7.24 & 0.70 & 0.53 $\pm$ 0.31 & 56810.87 $\pm$ 0.49 & 33.85 $\pm$ 10.69 & 0.88 $\pm$ 0.73 &   \\
305 & 84876 & -0.58477 & -2.42338 & 16.63 & 4.77 & 5.45 & 0.13 $\pm$ 0.10 & 56174.14 $\pm$ 0.42 & 21.99 $\pm$ 13.00 & 1.00 $\pm$ 0.87 &   \\
305 & 51513 & -0.93105 & -1.63498 & 14.52 & 6.56 & 1.90 & 0.00 $\pm$ 2.00 & 56414.19 $\pm$ 4.25 & 93.50 $\pm$ 39.03 & 1.00 $\pm$ 0.82 &   \\
305 & 12877 & -0.95509 & -2.10716 & 12.98 & 7.87 & 0.98 & 0.38 $\pm$ 0.18 & 56105.26 $\pm$ 0.38 & 20.09 $\pm$ 6.09 & 0.57 $\pm$ 0.36 &   \\
305 & 4548 & -0.56145 & -1.53921 & 14.58 & 6.51 & 1.15 & 0.51 $\pm$ 0.52 & 56756.77 $\pm$ 5.88 & 110.17 $\pm$ 64.66 & 1.00 $\pm$ 1.53 &   \\
305 & 10965 & -0.64853 & -2.47010 & 14.11 & 6.91 & 0.21 & 1.28 $\pm$ 2.00 & 56085.19 $\pm$ 8.33 & 39.80 $\pm$ 100.18 & 1.00 $\pm$ 8.99 &   \\
305 & 22105 & -0.58169 & -2.49046 & 14.89 & 6.24 & 1.53 & 0.20 $\pm$ 0.10 & 56118.50 $\pm$ 0.38 & 63.55 $\pm$ 23.34 & 0.40 $\pm$ 0.23 & DET. \\
305 & 24945 & -0.46515 & -2.49488 & 13.48 & 1.55 & 2.64 & 0.25 $\pm$ 0.06 & 56377.50 $\pm$ 0.13 & 24.20 $\pm$ 4.39 & 0.85 $\pm$ 0.27 & DET. \\
305 & 51557 & -0.40626 & -2.54378 & 15.11 & 1.31 & 1.72 & 0.00 $\pm$ 2.00 & 56517.13 $\pm$ 0.07 & 17.80 $\pm$ 5.87 & 0.81 $\pm$ 0.47 & DET. \\
305 & 87644 & -0.45480 & -2.61041 & 13.46 & 1.49 & 0.22 & 0.96 $\pm$ 1.90 & 55808.57 $\pm$ 2.37 & 55.80 $\pm$ 74.65 & 0.59 $\pm$ 2.37 &   \\
305 & 45586 & 0.04629 & -1.98380 & 14.73 & 0.93 & 0.85 & 0.57 $\pm$ 2.00 & 57262.22 $\pm$ 3.16 & 25.32 $\pm$ 296.84 & 1.00 $\pm$ 29.41 &   \\
305 & 76051 & -0.45942 & -2.21995 & 13.17 & 1.12 & 1.14 & 0.52 $\pm$ 0.41 & 56141.48 $\pm$ 0.91 & 66.26 $\pm$ 36.77 & 1.00 $\pm$ 1.22 & DET. \\
305 & 57836 & -0.13380 & -1.64958 & 14.72 & 1.25 & 0.79 & 0.51 $\pm$ 2.00 & 56739.29 $\pm$ 0.39 & 7.13 $\pm$ 315.72 & 1.00 $\pm$ 101.77 & DET. \\
305 & 79184 & -0.04161 & -1.69272 & 16.21 & 0.95 & 1.90 & 0.31 $\pm$ 0.58 & 56841.13 $\pm$ 0.48 & 5.47 $\pm$ 6.73 & 0.79 $\pm$ 1.94 & DET. \\
305 & 87146 & 0.09125 & -1.87919 & 13.12 & 1.10 & 0.39 & 0.58 $\pm$ 0.35 & 55799.12 $\pm$ 0.94 & 39.08 $\pm$ 14.99 & 0.44 $\pm$ 0.43 & DET. \\
305 & 78425 & 0.19782 & -1.69504 & 13.78 & 1.16 & 0.29 & 0.01 $\pm$ 0.05 & 56375.90 $\pm$ 0.43 & 166.71 $\pm$ 738.34 & 0.00 $\pm$ 0.02 &   \\
305 & 83270 & 0.16365 & -2.24116 & 14.99 & 1.01 & 1.93 & 0.00 $\pm$ 0.00 & 56370.22 $\pm$ 0.12 & 750.00 $\pm$ 1053.09 & 0.01 $\pm$ 0.01 & DET. \\
305 & 52035 & -0.30808 & -1.99248 & 16.27 & 0.82 & 3.59 & 0.14 $\pm$ 0.07 & 56170.20 $\pm$ 0.26 & 32.14 $\pm$ 10.55 & 0.62 $\pm$ 0.32 & DET. \\
305 & 96244 & 0.07972 & -2.07456 & 14.35 & 0.97 & 0.32 & 0.23 $\pm$ 1.68 & 56610.03 $\pm$ 10.80 & 67.30 $\pm$ 69.96 & 0.54 $\pm$ 1.70 & DET. \\
305 & 18925 & 0.16883 & -2.48300 & 14.59 & 1.24 & 0.88 & 0.58 $\pm$ 0.49 & 55800.69 $\pm$ 1.34 & 58.23 $\pm$ 31.57 & 1.00 $\pm$ 1.35 & DET. \\
319 & 21482 & 0.04500 & -1.39533 & 14.64 & 1.32 & 2.77 & 0.24 $\pm$ 0.37 & 56368.42 $\pm$ 0.13 & 10.38 $\pm$ 12.12 & 1.00 $\pm$ 1.87 &   \\
319 & 75363 & 0.13492 & -1.13231 & 14.40 & 1.65 & 3.02 & 0.24 $\pm$ 0.13 & 55808.19 $\pm$ 0.94 & 29.76 $\pm$ 8.42 & 1.00 $\pm$ 0.60 & DET. \\
319 & 81852 & 0.12729 & -1.14514 & 12.37 & 1.72 & 0.47 & 0.18 $\pm$ 0.17 & 55827.51 $\pm$ 1.92 & 124.93 $\pm$ 93.18 & 0.10 $\pm$ 0.11 & DET. \\
319 & 3050 & 0.16601 & -0.44743 & 13.79 & 4.88 & 11.59 & 0.03 $\pm$ 0.03 & 57245.48 $\pm$ 0.45 & 248.58 $\pm$ 174.11 & 0.47 $\pm$ 0.37 &   \\
319 & 19761 & 0.10893 & -0.49540 & 15.04 & 4.42 & 0.59 & 0.47 $\pm$ 1.65 & 56866.31 $\pm$ 1.03 & 22.11 $\pm$ 31.06 & 1.00 $\pm$ 4.00 &   \\
319 & 29885 & 0.06187 & -0.52356 & 13.78 & 4.65 & 2.07 & 0.00 $\pm$ 2.00 & 56469.09 $\pm$ 1.09 & 38.41 $\pm$ 8.88 & 0.53 $\pm$ 0.30 &   \\
319 & 47428 & -0.27338 & -1.07622 & 13.23 & 2.11 & 1.96 & 0.29 $\pm$ 0.05 & 56897.46 $\pm$ 0.54 & 62.39 $\pm$ 6.11 & 0.76 $\pm$ 0.16 & DET. \\
319 & 87262 & -0.28093 & -1.14837 & 13.82 & 1.61 & 2.08 & 0.16 $\pm$ 0.14 & 56522.87 $\pm$ 0.37 & 19.14 $\pm$ 6.81 & 0.75 $\pm$ 0.50 & DET. \\
319 & 95634 & 0.19390 & -1.35658 & 13.54 & 1.84 & 0.61 & 0.12 $\pm$ 0.30 & 55844.46 $\pm$ 1.74 & 54.69 $\pm$ 86.36 & 0.09 $\pm$ 0.24 & DET. \\
319 & 95942 & 0.06391 & -0.81765 & 16.34 & 1.52 & 3.21 & 0.00 $\pm$ 2.00 & 55391.28 $\pm$ 2.61 & 750.00 $\pm$ 11455.69 & 0.00 $\pm$ 0.07 &   \\
319 & 14507 & -0.29881 & -0.64897 & 13.64 & 3.36 & 3.34 & 0.11 $\pm$ 0.04 & 56851.57 $\pm$ 0.88 & 72.95 $\pm$ 20.02 & 0.43 $\pm$ 0.19 &   \\
319 & 82449 & -0.19826 & -0.77694 & 13.46 & 3.14 & 0.21 & 0.00 $\pm$ 2.00 & 56976.02 $\pm$ 12.07 & 61.53 $\pm$ 24.79 & 1.00 $\pm$ 1.26 &   \\
319 & 70390 & -0.03710 & -1.48369 & 12.91 & 1.31 & 0.98 & 0.55 $\pm$ 0.49 & 56822.26 $\pm$ 2.31 & 42.98 $\pm$ 23.92 & 1.00 $\pm$ 1.42 & DET. \\
319 & 53886 & -0.10963 & -0.90446 & 13.58 & 2.15 & 0.61 & 0.00 $\pm$ 2.00 & 56179.94 $\pm$ 0.99 & 20.58 $\pm$ 8.78 & 1.00 $\pm$ 0.89 &   \\
319 & 6989 & -0.93746 & -0.45560 & 14.67 & 6.43 & 1.73 & 0.00 $\pm$ 2.00 & 56184.61 $\pm$ 1.29 & 27.62 $\pm$ 9.47 & 0.96 $\pm$ 0.67 &   \\
319 & 79778 & -1.00800 & -0.95836 & 13.14 & 7.73 & 1.46 & 0.26 $\pm$ 0.09 & 56131.21 $\pm$ 0.28 & 8.93 $\pm$ 1.98 & 0.55 $\pm$ 0.26 &   \\
319 & 87060 & -0.93844 & -0.97172 & 12.95 & 7.89 & 0.46 & 0.00 $\pm$ 2.00 & 56861.64 $\pm$ 4.12 & 24.27 $\pm$ 14.89 & 0.23 $\pm$ 0.43 &   \\
319 & 3547 & -0.27303 & -0.44503 & 12.71 & 8.10 & 0.75 & 0.32 $\pm$ 0.24 & 57259.82 $\pm$ 0.42 & 36.72 $\pm$ 17.98 & 0.41 $\pm$ 0.39 &   \\
319 & 8718 & -0.54848 & -1.00606 & 13.45 & 7.47 & 1.76 & 0.28 $\pm$ 0.13 & 56507.59 $\pm$ 0.23 & 12.70 $\pm$ 4.02 & 0.65 $\pm$ 0.39 &   \\
319 & 95579 & -0.57219 & -1.16388 & 13.24 & 7.64 & 0.35 & 0.99 $\pm$ 1.13 & 56851.67 $\pm$ 1.15 & 52.72 $\pm$ 40.65 & 1.00 $\pm$ 2.36 &   \\
319 & 3978 & -0.50597 & -0.44617 & 16.77 & 4.64 & 3.49 & 0.06 $\pm$ 0.71 & 55359.28 $\pm$ 1.27 & 97.17 $\pm$ 155.95 & 1.00 $\pm$ 2.65 &   \\
319 & 49598 & -1.04837 & -1.08947 & 14.81 & 6.31 & 6.51 & 0.08 $\pm$ 0.05 & 56793.35 $\pm$ 0.49 & 48.94 $\pm$ 15.19 & 1.00 $\pm$ 0.45 &   \\
319 & 59552 & -0.91198 & -1.10962 & 12.93 & 7.91 & 0.43 & 0.84 $\pm$ 1.39 & 56431.84 $\pm$ 2.73 & 32.45 $\pm$ 29.84 & 1.00 $\pm$ 3.06 &   \\
319 & 68573 & -0.51615 & -0.75423 & 15.92 & 5.37 & 1.66 & 0.00 $\pm$ 2.00 & 56889.59 $\pm$ 1.98 & 24.28 $\pm$ 40.59 & 0.18 $\pm$ 0.43 &   \\
319 & 29850 & -1.02870 & -0.68188 & 15.19 & 5.99 & 1.58 & 0.08 $\pm$ 0.07 & 56186.79 $\pm$ 1.98 & 84.41 $\pm$ 35.71 & 0.22 $\pm$ 0.14 &   \\
319 & 38669 & -1.03987 & -0.69888 & 15.03 & 6.13 & 0.82 & 0.24 $\pm$ 0.32 & 56800.29 $\pm$ 2.19 & 42.95 $\pm$ 41.07 & 0.24 $\pm$ 0.41 &   \\
319 & 81917 & -0.91565 & -0.78509 & 13.94 & 7.05 & 1.67 & 0.39 $\pm$ 0.07 & 56812.45 $\pm$ 0.50 & 70.14 $\pm$ 8.45 & 1.00 $\pm$ 0.26 &   \\
319 & 19250 & -0.84143 & -1.02642 & 13.95 & 7.04 & 0.70 & 0.00 $\pm$ 2.00 & 56135.02 $\pm$ 0.07 & 6.90 $\pm$ 0.81 & 1.00 $\pm$ 0.22 &   \\
319 & 90727 & -0.78873 & -1.16034 & 13.77 & 7.19 & 0.23 & 1.19 $\pm$ 2.00 & 56441.84 $\pm$ 3.81 & 69.97 $\pm$ 115.95 & 1.00 $\pm$ 5.57 &   \\
319 & 38914 & -0.82286 & -1.42640 & 14.16 & 6.87 & 1.10 & 0.19 $\pm$ 0.17 & 56170.63 $\pm$ 0.83 & 50.60 $\pm$ 18.41 & 0.47 $\pm$ 0.33 &   \\
319 & 79079 & -0.85193 & -1.49861 & 16.69 & 4.71 & 4.95 & 0.00 $\pm$ 0.02 & 56887.66 $\pm$ 0.42 & 750.00 $\pm$ 8127.74 & 0.01 $\pm$ 0.09 & DET. \\
319 & 102314 & -0.86366 & -1.54145 & 13.55 & 7.38 & 0.30 & 0.91 $\pm$ 2.00 & 55813.18 $\pm$ 3.16 & 38.35 $\pm$ 243.61 & 0.72 $\pm$ 13.22 &   \\
319 & 5331 & -0.77071 & -0.81517 & 12.30 & 8.45 & 0.31 & 0.04 $\pm$ 0.07 & 56123.62 $\pm$ 1.15 & 49.76 $\pm$ 77.66 & 0.02 $\pm$ 0.03 &   \\
319 & 33050 & -0.71690 & -0.86786 & 13.65 & 7.30 & 0.37 & 0.94 $\pm$ 2.00 & 56153.75 $\pm$ 0.99 & 8.41 $\pm$ 12.59 & 1.00 $\pm$ 4.64 &   \\
319 & 80490 & -1.19889 & -0.96551 & 13.05 & 7.80 & 0.41 & 0.15 $\pm$ 0.05 & 56327.44 $\pm$ 3.55 & 387.16 $\pm$ 80.24 & 0.08 $\pm$ 0.03 &   \\
319 & 16915 & 0.08345 & -0.83682 & 12.57 & 2.96 & 3.47 & 0.15 $\pm$ 0.01 & 56137.23 $\pm$ 0.05 & 21.96 $\pm$ 1.22 & 0.86 $\pm$ 0.08 &   \\
319 & 30059 & 0.16936 & -0.86199 & 13.84 & 2.48 & 0.68 & 0.27 $\pm$ 0.28 & 55836.77 $\pm$ 0.76 & 24.08 $\pm$ 17.45 & 0.27 $\pm$ 0.35 &   \\
319 & 40061 & 0.18045 & -0.88065 & 12.28 & 3.04 & 0.52 & 0.16 $\pm$ 0.48 & 56551.22 $\pm$ 0.89 & 18.57 $\pm$ 8.75 & 0.38 $\pm$ 0.42 &   \\
319 & 55195 & 0.15508 & -0.90955 & 13.84 & 2.20 & 0.43 & 0.81 $\pm$ 1.72 & 56498.58 $\pm$ 0.84 & 19.24 $\pm$ 23.76 & 1.00 $\pm$ 3.82 &   \\
319 & 55412 & 0.04249 & -0.90998 & 14.34 & 1.71 & 0.73 & 0.00 $\pm$ 2.00 & 55364.21 $\pm$ 2.28 & 29.03 $\pm$ 20.84 & 1.00 $\pm$ 1.49 &   \\
319 & 1611 & -0.18819 & -1.35974 & 16.19 & 0.79 & 13.23 & 0.00 $\pm$ 0.01 & 55388.41 $\pm$ 0.18 & 750.00 $\pm$ 2191.94 & 0.05 $\pm$ 0.16 &   \\
319 & 3488 & -0.22720 & -1.36323 & 16.36 & 4.96 & 6.52 & 0.00 $\pm$ 2.00 & 55398.73 $\pm$ 3.47 & 44.95 $\pm$ 72.41 & 1.00 $\pm$ 2.41 &   \\
319 & 16104 & -0.14236 & -1.38481 & 14.40 & 1.50 & 1.04 & 0.42 $\pm$ 0.31 & 56071.07 $\pm$ 3.13 & 58.45 $\pm$ 22.35 & 1.00 $\pm$ 0.97 & DET. \\
319 & 38154 & -0.29430 & -1.42327 & 14.32 & 1.24 & 0.45 & 0.55 $\pm$ 2.00 & 55366.95 $\pm$ 3.14 & 24.10 $\pm$ 45.29 & 1.00 $\pm$ 5.66 &   \\
319 & 101943 & 0.00588 & -0.81570 & 16.62 & 1.38 & 5.06 & 0.00 $\pm$ 2.00 & 55377.47 $\pm$ 6.32 & 750.00 $\pm$ 2627.19 & 0.05 $\pm$ 0.19 &   \\
319 & 51595 & -0.43142 & -0.71629 & 14.27 & 3.27 & 1.63 & 0.39 $\pm$ 0.70 & 56488.37 $\pm$ 2.06 & 14.97 $\pm$ 16.18 & 1.00 $\pm$ 2.60 &   \\
319 & 70966 & -0.39831 & -0.75228 & 14.90 & 6.24 & 2.05 & 0.18 $\pm$ 0.12 & 56792.55 $\pm$ 0.49 & 37.61 $\pm$ 13.54 & 0.61 $\pm$ 0.38 &   \\
319 & 87886 & -0.39165 & -0.78417 & 15.28 & 1.99 & 1.39 & 0.00 $\pm$ 2.00 & 56889.73 $\pm$ 1.26 & 21.75 $\pm$ 22.36 & 0.19 $\pm$ 0.29 &   \\
319 & 102369 & -0.56143 & -1.53920 & 14.61 & 6.48 & 1.21 & 0.24 $\pm$ 0.07 & 56758.03 $\pm$ 2.01 & 191.44 $\pm$ 34.05 & 0.37 $\pm$ 0.12 & DET. \\
319 & 10859 & -0.75308 & -0.64291 & 15.51 & 5.72 & 6.45 & 0.05 $\pm$ 0.04 & 56189.52 $\pm$ 2.18 & 44.61 $\pm$ 20.37 & 1.00 $\pm$ 0.77 &   \\
319 & 27112 & 0.04135 & -1.03979 & 13.94 & 1.65 & 1.97 & 0.23 $\pm$ 0.52 & 55399.88 $\pm$ 1.52 & 29.47 $\pm$ 32.79 & 1.00 $\pm$ 2.13 &   \\
319 & 11870 & -0.34414 & -0.46156 & 14.70 & 6.40 & 3.87 & 0.00 $\pm$ 2.00 & 56566.44 $\pm$ 0.06 & 38.23 $\pm$ 5.63 & 0.66 $\pm$ 0.15 &   \\
319 & 63316 & -0.36026 & -0.56140 & 14.98 & 6.17 & 0.89 & 0.24 $\pm$ 0.13 & 56871.16 $\pm$ 3.41 & 189.97 $\pm$ 75.60 & 0.27 $\pm$ 0.19 &   \\
319 & 57363 & -0.74284 & -0.55253 & 14.33 & 6.72 & 0.42 & 0.93 $\pm$ 2.00 & 55847.21 $\pm$ 18.51 & 70.51 $\pm$ 158.91 & 1.00 $\pm$ 7.49 &   \\
319 & 8202 & -1.05960 & -0.45776 & 14.08 & 6.93 & 2.18 & 0.21 $\pm$ 0.36 & 55368.37 $\pm$ 2.12 & 64.86 $\pm$ 43.11 & 1.00 $\pm$ 1.35 &   \\
319 & 11838 & -1.11256 & -0.46606 & 13.16 & 7.72 & 0.30 & 0.31 $\pm$ 0.32 & 56626.80 $\pm$ 5.04 & 112.49 $\pm$ 43.93 & 0.25 $\pm$ 0.24 &   \\
319 & 12120 & -1.06012 & -0.46627 & 14.25 & 6.79 & 0.94 & 0.60 $\pm$ 1.17 & 55373.41 $\pm$ 3.28 & 64.04 $\pm$ 79.50 & 1.00 $\pm$ 3.15 &   \\
319 & 17191 & -1.20267 & -0.47822 & 14.58 & 6.51 & 1.22 & 0.31 $\pm$ 0.26 & 57211.60 $\pm$ 9.31 & 68.93 $\pm$ 32.93 & 0.68 $\pm$ 0.85 &   \\
319 & 27358 & -1.18493 & -0.50014 & 11.83 & 8.84 & 0.81 & 0.53 $\pm$ 0.37 & 56377.24 $\pm$ 1.78 & 38.47 $\pm$ 17.20 & 0.75 $\pm$ 0.84 &   \\
347 & 27705 & -1.22849 & 1.32329 & 13.56 & 7.37 & 0.43 & 0.89 $\pm$ 1.33 & 57251.02 $\pm$ 0.49 & 9.90 $\pm$ 9.76 & 1.00 $\pm$ 2.96 & DET. \\
347 & 65904 & -0.82823 & 1.24752 & 13.74 & 7.22 & 2.97 & 0.15 $\pm$ 0.05 & 56547.31 $\pm$ 0.08 & 10.51 $\pm$ 1.69 & 1.00 $\pm$ 0.29 &   \\
347 & 69960 & -0.78207 & 1.23871 & 13.65 & 7.30 & 0.55 & 0.82 $\pm$ 0.86 & 56141.68 $\pm$ 0.65 & 13.68 $\pm$ 9.16 & 1.00 $\pm$ 2.00 &   \\
347 & 28677 & -0.86148 & 1.13687 & 15.16 & 6.02 & 1.34 & 0.46 $\pm$ 0.32 & 56138.18 $\pm$ 0.58 & 19.75 $\pm$ 8.39 & 1.00 $\pm$ 1.00 &   \\
347 & 44270 & -0.85012 & 1.10457 & 11.90 & 8.78 & 0.39 & 0.98 $\pm$ 1.65 & 56518.70 $\pm$ 1.66 & 53.48 $\pm$ 61.50 & 1.00 $\pm$ 3.49 &   \\
347 & 50357 & -0.70073 & 1.09192 & 15.07 & 6.09 & 1.33 & 0.26 $\pm$ 0.12 & 56134.30 $\pm$ 0.56 & 55.50 $\pm$ 17.96 & 0.46 $\pm$ 0.27 &   \\
347 & 90728 & -0.45960 & 1.57430 & 13.44 & 7.47 & 0.68 & 0.57 $\pm$ 0.42 & 56099.64 $\pm$ 1.03 & 29.82 $\pm$ 14.49 & 0.70 $\pm$ 0.83 & DET. \\
347 & 69362 & -1.18526 & 1.41394 & 14.21 & 6.82 & 2.82 & 0.08 $\pm$ 0.07 & 56822.72 $\pm$ 0.11 & 14.27 $\pm$ 2.21 & 0.92 $\pm$ 0.25 & DET. \\
347 & 50028 & -1.21774 & 0.90168 & 14.62 & 6.48 & 0.59 & 0.24 $\pm$ 0.27 & 57248.58 $\pm$ 0.30 & 9.74 $\pm$ 7.89 & 0.19 $\pm$ 0.25 &   \\
347 & 45377 & -0.78496 & 1.47771 & 13.80 & 7.17 & 0.54 & 0.47 $\pm$ 0.50 & 56867.17 $\pm$ 1.10 & 39.89 $\pm$ 22.39 & 0.62 $\pm$ 0.87 & DET. \\
347 & 41875 & -0.74337 & 0.91837 & 15.42 & 5.79 & 2.31 & 0.05 $\pm$ 0.11 & 56062.46 $\pm$ 1.56 & 129.53 $\pm$ 32.61 & 0.62 $\pm$ 0.25 &   \\
347 & 49702 & -0.70991 & 0.89991 & 13.72 & 7.24 & 6.11 & 0.12 $\pm$ 0.06 & 56169.68 $\pm$ 0.46 & 22.39 $\pm$ 5.32 & 1.00 $\pm$ 0.45 &   \\
347 & 22839 & -0.90341 & 1.51525 & 15.17 & 6.01 & 1.29 & 0.06 $\pm$ 0.10 & 56834.96 $\pm$ 1.03 & 236.52 $\pm$ 375.49 & 0.07 $\pm$ 0.13 & DET. \\
347 & 38514 & -0.94772 & 0.92390 & 13.67 & 7.28 & 0.78 & 0.52 $\pm$ 0.17 & 56554.34 $\pm$ 0.31 & 17.30 $\pm$ 3.64 & 0.68 $\pm$ 0.35 &   \\
347 & 48421 & -1.03510 & 0.90071 & 15.21 & 5.97 & 0.98 & 0.54 $\pm$ 0.78 & 57259.26 $\pm$ 0.93 & 21.71 $\pm$ 21.64 & 1.00 $\pm$ 2.28 &   \\
347 & 16027 & -0.66306 & 0.97723 & 14.37 & 6.68 & 0.70 & 0.67 $\pm$ 1.72 & 56205.40 $\pm$ 1.78 & 10.70 $\pm$ 13.77 & 1.00 $\pm$ 4.29 &   \\
347 & 25797 & -1.04459 & 1.14102 & 13.90 & 7.08 & 2.60 & 0.29 $\pm$ 0.04 & 56167.85 $\pm$ 0.17 & 34.97 $\pm$ 2.94 & 1.00 $\pm$ 0.16 &   \\
347 & 34396 & -1.01718 & 1.12279 & 14.61 & 6.48 & 1.76 & 0.28 $\pm$ 0.10 & 56134.96 $\pm$ 0.27 & 28.61 $\pm$ 6.80 & 0.72 $\pm$ 0.33 &   \\
347 & 5981 & -0.56564 & 1.73438 & 14.07 & 6.94 & 8.91 & 0.10 $\pm$ 0.01 & 56102.96 $\pm$ 0.02 & 14.14 $\pm$ 1.28 & 1.00 $\pm$ 0.14 & DET. \\
347 & 82999 & -0.56626 & 1.02288 & 15.83 & 5.45 & 6.94 & 0.08 $\pm$ 0.04 & 56536.81 $\pm$ 0.13 & 34.36 $\pm$ 9.64 & 1.00 $\pm$ 0.38 &   \\
347 & 40812 & -0.65045 & 1.30091 & 13.09 & 7.77 & 1.00 & 0.57 $\pm$ 0.16 & 56516.59 $\pm$ 1.43 & 114.85 $\pm$ 21.24 & 1.00 $\pm$ 0.45 &   \\
347 & 41193 & 0.03728 & 1.29910 & 15.32 & 5.88 & 1.81 & 0.39 $\pm$ 0.66 & 56853.05 $\pm$ 3.97 & 57.45 $\pm$ 65.74 & 1.00 $\pm$ 2.41 &   \\
347 & 69720 & -0.04291 & 0.68396 & 14.99 & 3.34 & 3.43 & 0.12 $\pm$ 0.04 & 56576.07 $\pm$ 0.32 & 34.10 $\pm$ 8.66 & 0.52 $\pm$ 0.19 &   \\
347 & 92482 & 0.03445 & 0.63426 & 14.13 & 6.89 & 5.13 & 0.15 $\pm$ 0.10 & 57251.00 $\pm$ 0.30 & 34.26 $\pm$ 17.98 & 0.92 $\pm$ 0.70 &   \\
347 & 35068 & -0.50154 & 1.12892 & 14.80 & 1.82 & 0.39 & 0.99 $\pm$ 2.00 & 56178.09 $\pm$ 7.67 & 90.75 $\pm$ 124.03 & 1.00 $\pm$ 4.27 &   \\
347 & 72620 & -0.37091 & 1.05774 & 16.94 & 4.50 & 3.60 & 0.18 $\pm$ 0.36 & 57243.78 $\pm$ 0.71 & 13.38 $\pm$ 14.20 & 1.00 $\pm$ 1.95 &   \\
347 & 28089 & -0.42203 & 1.50986 & 13.73 & 7.23 & 0.34 & 0.25 $\pm$ 0.43 & 57248.53 $\pm$ 0.31 & 5.70 $\pm$ 6.57 & 0.12 $\pm$ 0.24 &   \\
347 & 62565 & -0.40420 & 1.44447 & 12.85 & 1.74 & 0.29 & 0.59 $\pm$ 0.52 & 56113.40 $\pm$ 0.21 & 6.82 $\pm$ 3.57 & 0.34 $\pm$ 0.48 &   \\
347 & 83068 & -0.39933 & 1.40547 & 14.04 & 1.16 & 0.59 & 0.01 $\pm$ 0.02 & 56115.01 $\pm$ 0.28 & 324.86 $\pm$ 957.76 & 0.00 $\pm$ 0.01 &   \\
347 & 3785 & -0.42851 & 1.00270 & 16.14 & 1.92 & 1.73 & 0.36 $\pm$ 0.70 & 56136.46 $\pm$ 1.90 & 26.14 $\pm$ 31.35 & 1.00 $\pm$ 2.56 &   \\
347 & 18360 & -0.05394 & 1.52645 & 15.10 & 6.07 & 0.57 & 0.09 $\pm$ 0.07 & 56827.30 $\pm$ 3.37 & 369.03 $\pm$ 206.61 & 0.06 $\pm$ 0.04 &   \\
347 & 21146 & 0.04363 & 1.52078 & 14.73 & 6.38 & 0.64 & 0.55 $\pm$ 1.60 & 57249.48 $\pm$ 0.23 & 4.79 $\pm$ 8.10 & 0.76 $\pm$ 3.35 &   \\
347 & 61846 & -0.07325 & 1.44093 & 14.96 & 1.48 & 0.71 & 0.41 $\pm$ 0.86 & 56202.00 $\pm$ 1.53 & 19.65 $\pm$ 27.17 & 0.42 $\pm$ 1.24 &   \\
347 & 75581 & -0.01650 & 1.41358 & 14.89 & 6.24 & 4.85 & 0.08 $\pm$ 0.02 & 56212.16 $\pm$ 1.90 & 117.91 $\pm$ 20.14 & 0.47 $\pm$ 0.14 &   \\
347 & 75845 & -0.04066 & 1.41306 & 13.91 & 1.43 & 5.01 & 0.13 $\pm$ 0.02 & 56123.79 $\pm$ 0.02 & 3.89 $\pm$ 0.41 & 1.00 $\pm$ 0.16 &   \\
347 & 21402 & 0.00616 & 0.96843 & 13.20 & 1.75 & 0.92 & 0.54 $\pm$ 2.00 & 56498.69 $\pm$ 4.68 & 6.81 $\pm$ 20.76 & 1.00 $\pm$ 10.10 &   \\
347 & 68700 & 0.03206 & 0.87129 & 13.70 & 7.26 & 0.82 & 0.61 $\pm$ 0.98 & 57247.63 $\pm$ 0.20 & 3.59 $\pm$ 3.63 & 1.00 $\pm$ 2.62 &   \\
347 & 39402 & -0.22791 & 0.93399 & 13.43 & 7.49 & 0.17 & 0.18 $\pm$ 0.14 & 56115.49 $\pm$ 0.88 & 61.17 $\pm$ 35.99 & 0.04 $\pm$ 0.03 &   \\
347 & 72106 & -0.27629 & 0.86957 & 13.73 & 1.49 & 0.65 & 0.49 $\pm$ 2.00 & 55807.67 $\pm$ 0.89 & 16.18 $\pm$ 31.14 & 1.00 $\pm$ 5.27 &   \\
347 & 51199 & 0.16339 & 1.46328 & 14.64 & 1.63 & 2.52 & 0.29 $\pm$ 0.05 & 56101.02 $\pm$ 0.28 & 47.87 $\pm$ 5.45 & 1.00 $\pm$ 0.21 &   \\
347 & 44668 & 0.19562 & 0.92445 & 12.67 & 1.89 & 0.34 & 0.34 $\pm$ 1.88 & 56733.61 $\pm$ 8.45 & 79.68 $\pm$ 84.73 & 1.00 $\pm$ 3.43 &   \\
347 & 57724 & 0.06212 & 0.89838 & 14.33 & 6.72 & 1.66 & 0.28 $\pm$ 0.25 & 57239.30 $\pm$ 0.15 & 10.35 $\pm$ 4.66 & 1.00 $\pm$ 0.92 &   \\
347 & 87437 & 0.12963 & 0.83843 & 14.69 & 2.21 & 0.90 & 0.61 $\pm$ 0.56 & 56566.02 $\pm$ 1.28 & 36.40 $\pm$ 20.83 & 1.00 $\pm$ 1.48 &   \\
347 & 7923 & -0.27876 & 1.17927 & 13.65 & 7.30 & 1.45 & 0.10 $\pm$ 0.44 & 56831.99 $\pm$ 0.53 & 8.30 $\pm$ 10.91 & 0.31 $\pm$ 0.72 &   \\
347 & 76798 & 0.11414 & 1.58643 & 13.53 & 1.56 & 0.31 & 1.08 $\pm$ 2.00 & 56188.78 $\pm$ 2.18 & 28.44 $\pm$ 38.56 & 1.00 $\pm$ 4.47 &   \\
347 & 50518 & 0.12593 & 1.09631 & 11.86 & 8.82 & 0.81 & 0.68 $\pm$ 0.50 & 55877.24 $\pm$ 15.06 & 333.28 $\pm$ 143.49 & 1.00 $\pm$ 1.25 &   \\
347 & 35700 & -0.23557 & 1.31220 & 13.04 & 1.75 & 1.33 & 0.23 $\pm$ 0.03 & 56208.96 $\pm$ 1.14 & 75.24 $\pm$ 6.34 & 0.40 $\pm$ 0.07 &   \\
347 & 39564 & -0.27058 & 1.30506 & 12.42 & 1.83 & 0.23 & 0.84 $\pm$ 2.00 & 56183.67 $\pm$ 0.16 & 3.91 $\pm$ 6.68 & 1.00 $\pm$ 5.66 &   \\
347 & 82357 & -0.17561 & 1.22425 & 13.15 & 1.99 & 0.94 & 0.43 $\pm$ 0.14 & 56023.72 $\pm$ 1.60 & 57.66 $\pm$ 10.67 & 0.82 $\pm$ 0.38 &   \\
347 & 94666 & -0.24569 & 0.63450 & 16.10 & 5.21 & 3.32 & 0.02 $\pm$ 0.02 & 57248.37 $\pm$ 0.93 & 750.00 $\pm$ 589.38 & 0.07 $\pm$ 0.06 &   \\
347 & 18361 & 0.17027 & 0.79111 & 13.05 & 2.77 & 0.41 & 0.91 $\pm$ 1.11 & 56201.32 $\pm$ 1.82 & 22.43 $\pm$ 16.96 & 1.00 $\pm$ 2.42 &   \\
347 & 51805 & 0.11390 & 0.72254 & 14.13 & 2.60 & 2.11 & 0.27 $\pm$ 0.25 & 55821.68 $\pm$ 1.91 & 89.87 $\pm$ 63.45 & 0.64 $\pm$ 0.74 &   \\
347 & 52306 & 0.11364 & 0.72152 & 14.90 & 2.59 & 1.41 & 0.06 $\pm$ 0.13 & 55815.24 $\pm$ 2.27 & 124.77 $\pm$ 236.99 & 0.09 $\pm$ 0.20 &   \\
347 & 52578 & 0.11422 & 0.72098 & 14.19 & 2.47 & 0.45 & 0.75 $\pm$ 2.00 & 55808.24 $\pm$ 1.60 & 20.87 $\pm$ 86.28 & 1.00 $\pm$ 11.65 &   \\
347 & 76171 & 0.06529 & 0.67171 & 14.44 & 3.06 & 0.91 & 0.47 $\pm$ 0.26 & 56534.89 $\pm$ 0.46 & 27.08 $\pm$ 9.37 & 0.73 $\pm$ 0.59 &   \\
347 & 85434 & 0.13499 & 0.65202 & 15.06 & 6.10 & 1.95 & 0.27 $\pm$ 0.24 & 57253.81 $\pm$ 0.34 & 11.93 $\pm$ 5.88 & 1.00 $\pm$ 0.98 &   \\
361 & 56975 & -1.16291 & 2.35824 & 12.35 & 8.40 & 0.34 & 0.00 $\pm$ 2.00 & 56878.28 $\pm$ 1.08 & 36.32 $\pm$ 7.67 & 0.28 $\pm$ 0.12 & DET. \\
361 & 76350 & -1.16879 & 2.32008 & 13.48 & 7.44 & 0.25 & 0.07 $\pm$ 0.07 & 56557.69 $\pm$ 0.77 & 39.47 $\pm$ 30.03 & 0.02 $\pm$ 0.02 &   \\
361 & 1531 & -1.12199 & 1.91578 & 16.46 & 4.91 & 4.79 & 0.00 $\pm$ 2.00 & 56000.07 $\pm$ 0.42 & 28.13 $\pm$ 38.00 & 1.00 $\pm$ 2.07 &   \\
361 & 41821 & -1.09120 & 2.75287 & 12.05 & 8.66 & 0.22 & 1.19 $\pm$ 2.00 & 55995.26 $\pm$ 1.14 & 16.35 $\pm$ 77.04 & 1.00 $\pm$ 15.11 &   \\
361 & 12561 & -1.09206 & 2.26178 & 14.45 & 6.62 & 1.47 & 0.42 $\pm$ 0.10 & 56805.57 $\pm$ 0.18 & 20.36 $\pm$ 3.27 & 0.93 $\pm$ 0.33 & DET. \\
361 & 10285 & -0.73004 & 2.26764 & 13.78 & 7.18 & 0.14 & 1.41 $\pm$ 2.00 & 55996.67 $\pm$ 1.16 & 9.71 $\pm$ 95.03 & 1.00 $\pm$ 34.25 &   \\
361 & 52033 & -0.75845 & 2.19113 & 13.46 & 7.46 & 0.19 & 0.22 $\pm$ 0.34 & 55826.42 $\pm$ 2.38 & 73.90 $\pm$ 89.76 & 0.05 $\pm$ 0.09 &   \\
361 & 97832 & -0.47548 & 2.65704 & 17.19 & 4.28 & 4.43 & 0.00 $\pm$ 2.00 & 56002.03 $\pm$ 0.64 & 75.50 $\pm$ 86.87 & 0.20 $\pm$ 0.27 &   \\
361 & 5197 & -1.05840 & 2.09186 & 13.66 & 7.29 & 0.23 & 0.02 $\pm$ 0.03 & 55998.29 $\pm$ 0.81 & 298.13 $\pm$ 376.61 & 0.01 $\pm$ 0.01 &   \\
361 & 38445 & -1.07630 & 2.02557 & 14.27 & 6.77 & 0.39 & 0.00 $\pm$ 2.00 & 55998.25 $\pm$ 1.02 & 6.25 $\pm$ 34.42 & 1.00 $\pm$ 18.12 &   \\
361 & 21718 & -0.80340 & 2.06324 & 17.30 & 4.19 & 9.42 & 0.00 $\pm$ 2.00 & 57230.88 $\pm$ 2.14 & 53.12 $\pm$ 82.17 & 1.00 $\pm$ 2.31 &   \\
361 & 62342 & -0.68267 & 1.99040 & 13.66 & 7.29 & 0.29 & 1.05 $\pm$ 2.00 & 55993.98 $\pm$ 1.06 & 11.16 $\pm$ 29.41 & 1.00 $\pm$ 8.24 &   \\
361 & 87414 & -0.78409 & 1.94430 & 13.57 & 7.36 & 2.69 & 0.00 $\pm$ 2.00 & 56559.44 $\pm$ 0.46 & 27.60 $\pm$ 14.54 & 0.37 $\pm$ 0.27 & DET. \\
361 & 14405 & -0.93416 & 2.62554 & 16.02 & 5.29 & 5.43 & 0.05 $\pm$ 0.07 & 56177.45 $\pm$ 1.50 & 23.03 $\pm$ 7.58 & 0.76 $\pm$ 0.51 &   \\
361 & 21506 & -0.87720 & 2.61239 & 13.31 & 7.59 & 0.32 & 0.00 $\pm$ 2.00 & 55998.95 $\pm$ 0.81 & 9.90 $\pm$ 135.11 & 0.39 $\pm$ 14.43 &   \\
361 & 4300 & -1.03976 & 2.27743 & 13.61 & 7.33 & 0.41 & 0.90 $\pm$ 2.00 & 55993.34 $\pm$ 1.28 & 18.76 $\pm$ 48.60 & 1.00 $\pm$ 7.35 &   \\
361 & 53577 & -0.88740 & 2.18240 & 16.58 & 4.81 & 1.22 & 0.50 $\pm$ 0.77 & 56126.72 $\pm$ 1.16 & 15.13 $\pm$ 14.75 & 1.00 $\pm$ 2.35 &   \\
361 & 43137 & -0.51274 & 2.75734 & 15.18 & 6.00 & 1.14 & 0.50 $\pm$ 0.40 & 56837.67 $\pm$ 0.86 & 26.41 $\pm$ 13.66 & 1.00 $\pm$ 1.20 &   \\
361 & 50705 & -0.60138 & 2.19309 & 14.38 & 6.68 & 1.23 & 0.09 $\pm$ 0.05 & 57169.25 $\pm$ 2.20 & 61.30 $\pm$ 26.57 & 0.16 $\pm$ 0.12 & DET. \\
361 & 15874 & -0.95946 & 2.44027 & 13.23 & 7.65 & 0.26 & 1.03 $\pm$ 1.90 & 56758.88 $\pm$ 4.13 & 47.75 $\pm$ 54.17 & 0.88 $\pm$ 3.34 & DET. \\
361 & 22926 & -0.95717 & 2.42706 & 13.48 & 7.45 & 0.50 & 0.80 $\pm$ 1.19 & 57239.72 $\pm$ 0.45 & 12.76 $\pm$ 12.12 & 1.00 $\pm$ 2.77 & DET. \\
361 & 21529 & -0.91129 & 1.87798 & 15.80 & 5.47 & 1.15 & 0.50 $\pm$ 2.00 & 56007.24 $\pm$ 2.90 & 21.22 $\pm$ 103.87 & 1.00 $\pm$ 11.86 &   \\
361 & 12353 & -0.68595 & 2.44860 & 16.73 & 4.68 & 2.98 & 0.26 $\pm$ 0.26 & 56202.52 $\pm$ 2.38 & 55.58 $\pm$ 39.73 & 0.80 $\pm$ 0.99 &   \\
361 & 98676 & -0.66323 & 2.28986 & 15.67 & 5.58 & 7.70 & 0.07 $\pm$ 0.02 & 56152.77 $\pm$ 0.10 & 37.47 $\pm$ 7.00 & 0.56 $\pm$ 0.14 &   \\
361 & 18788 & -0.46625 & 2.43707 & 16.99 & 1.39 & 2.49 & 0.23 $\pm$ 0.14 & 56140.04 $\pm$ 0.50 & 28.63 $\pm$ 12.22 & 0.69 $\pm$ 0.51 &   \\
361 & 32613 & -0.33669 & 1.85856 & 14.27 & 1.57 & 0.58 & 0.69 $\pm$ 0.46 & 56139.12 $\pm$ 0.85 & 35.84 $\pm$ 15.06 & 0.82 $\pm$ 0.95 &   \\
361 & 22795 & -0.01716 & 1.87776 & 12.67 & 1.50 & 3.15 & 0.08 $\pm$ 0.03 & 56193.99 $\pm$ 0.04 & 17.47 $\pm$ 1.38 & 0.97 $\pm$ 0.13 &   \\
361 & 46457 & -0.40727 & 2.20140 & 13.97 & 1.41 & 0.26 & 0.93 $\pm$ 2.00 & 57244.00 $\pm$ 0.48 & 8.67 $\pm$ 36.59 & 1.00 $\pm$ 14.21 &   \\
361 & 18865 & -0.04031 & 2.25114 & 14.23 & 1.72 & 0.51 & 0.00 $\pm$ 2.00 & 56770.51 $\pm$ 2.48 & 24.43 $\pm$ 15.77 & 1.00 $\pm$ 1.53 &   \\
361 & 27218 & -0.10863 & 2.23468 & 14.47 & 1.62 & 0.52 & 0.84 $\pm$ 2.00 & 56822.08 $\pm$ 1.98 & 6.10 $\pm$ 22.86 & 1.00 $\pm$ 12.09 &   \\
361 & 43970 & 0.01860 & 2.20288 & 16.77 & 1.31 & 7.08 & 0.00 $\pm$ 2.00 & 57234.07 $\pm$ 3.84 & 750.00 $\pm$ 2430.72 & 0.02 $\pm$ 0.06 &   \\
361 & 10581 & -0.40728 & 2.63586 & 13.36 & 1.29 & 0.21 & 1.27 $\pm$ 2.00 & 56798.55 $\pm$ 4.78 & 10.08 $\pm$ 30.46 & 1.00 $\pm$ 11.43 &   \\
361 & 8362 & -0.40087 & 2.08736 & 14.29 & 1.57 & 4.13 & 0.09 $\pm$ 0.08 & 56577.86 $\pm$ 1.45 & 45.75 $\pm$ 15.81 & 0.46 $\pm$ 0.30 &   \\
361 & 49925 & -0.32015 & 2.01253 & 13.88 & 1.73 & 2.10 & 0.33 $\pm$ 0.08 & 56828.54 $\pm$ 0.09 & 18.57 $\pm$ 2.75 & 1.00 $\pm$ 0.31 & DET. \\
361 & 15164 & -0.09392 & 2.07512 & 12.25 & 1.70 & 1.06 & 0.53 $\pm$ 0.09 & 56006.97 $\pm$ 1.07 & 102.38 $\pm$ 10.42 & 1.00 $\pm$ 0.27 &   \\
361 & 75500 & -0.00028 & 1.95998 & 14.89 & 1.37 & 0.79 & 0.00 $\pm$ 2.00 & 55769.35 $\pm$ 2.68 & 34.06 $\pm$ 9.86 & 1.00 $\pm$ 0.67 &   \\
361 & 80391 & 0.00243 & 1.95050 & 10.96 & 1.69 & 2.59 & 0.00 $\pm$ 2.00 & 56120.01 $\pm$ 0.36 & 93.04 $\pm$ 115.93 & 0.10 $\pm$ 0.15 &   \\
361 & 67302 & -0.15211 & 2.16101 & 16.21 & 1.41 & 1.98 & 0.00 $\pm$ 2.00 & 56816.24 $\pm$ 0.99 & 10.87 $\pm$ 8.17 & 1.00 $\pm$ 1.57 &   \\
361 & 21040 & -0.20217 & 1.88149 & 13.91 & 1.44 & 1.23 & 0.49 $\pm$ 0.52 & 56567.46 $\pm$ 0.38 & 12.92 $\pm$ 8.65 & 1.00 $\pm$ 1.63 &   \\
361 & 106680 & -0.14401 & 1.72347 & 15.64 & 1.40 & 1.83 & 0.32 $\pm$ 0.50 & 56541.75 $\pm$ 1.12 & 18.52 $\pm$ 16.22 & 1.00 $\pm$ 1.87 &   \\
361 & 58222 & 0.20321 & 1.81015 & 13.48 & 1.37 & 0.96 & 0.25 $\pm$ 0.10 & 56063.92 $\pm$ 1.06 & 67.04 $\pm$ 11.04 & 0.60 $\pm$ 0.22 &   \\
361 & 87882 & 0.22687 & 1.75299 & 13.49 & 1.39 & 0.59 & 0.00 $\pm$ 2.00 & 57211.88 $\pm$ 10.06 & 54.05 $\pm$ 124.83 & 0.06 $\pm$ 0.25 & DET. \\
375 & 23417 & -1.18272 & 3.33098 & 15.83 & 5.44 & 4.44 & 0.05 $\pm$ 0.03 & 56145.04 $\pm$ 0.28 & 59.65 $\pm$ 18.05 & 0.41 $\pm$ 0.16 &   \\
375 & 18930 & -0.76669 & 3.89112 & 11.69 & 8.96 & 0.92 & 0.00 $\pm$ 2.00 & 56194.06 $\pm$ 0.77 & 5.78 $\pm$ 11.47 & 1.00 $\pm$ 5.29 &   \\
375 & 58388 & -0.88437 & 3.62696 & 13.45 & 7.47 & 1.60 & 0.41 $\pm$ 0.14 & 56081.53 $\pm$ 0.45 & 17.08 $\pm$ 2.93 & 1.00 $\pm$ 0.49 & DET. \\
375 & 43564 & -0.48704 & 3.11488 & 16.80 & 4.62 & 2.22 & 0.32 $\pm$ 0.37 & 56188.17 $\pm$ 0.41 & 8.55 $\pm$ 6.42 & 1.00 $\pm$ 1.55 &   \\
375 & 16941 & -0.86037 & 2.97719 & 13.95 & 7.04 & 0.36 & 0.53 $\pm$ 0.93 & 56562.73 $\pm$ 0.53 & 11.85 $\pm$ 11.15 & 0.46 $\pm$ 1.13 & DET. \\
375 & 86731 & -0.86845 & 2.82911 & 14.50 & 6.58 & 4.01 & 0.20 $\pm$ 0.02 & 56839.20 $\pm$ 0.10 & 50.83 $\pm$ 3.16 & 0.99 $\pm$ 0.10 & DET. \\
375 & 79774 & -0.65186 & 3.41172 & 13.98 & 7.02 & 0.20 & 1.24 $\pm$ 2.00 & 56789.67 $\pm$ 4.09 & 36.73 $\pm$ 132.14 & 1.00 $\pm$ 12.20 & DET. \\
375 & 8388 & -0.09215 & 3.36281 & 17.25 & 4.24 & 1.68 & 0.21 $\pm$ 0.52 & 56113.79 $\pm$ 1.01 & 21.02 $\pm$ 18.98 & 1.00 $\pm$ 1.83 &   \\
375 & 42857 & -0.48704 & 3.11487 & 16.96 & 4.48 & 2.76 & 0.28 $\pm$ 0.25 & 56188.07 $\pm$ 0.47 & 13.80 $\pm$ 8.56 & 1.00 $\pm$ 1.17 &   \\
375 & 23711 & 0.06360 & 3.69847 & 13.28 & 7.61 & 0.93 & 0.00 $\pm$ 2.00 & 55988.93 $\pm$ 1.05 & 65.53 $\pm$ 280.05 & 0.03 $\pm$ 0.16 &   \\
375 & 72168 & 0.24988 & 3.05546 & 16.72 & 4.69 & 5.05 & 0.00 $\pm$ 2.00 & 55995.52 $\pm$ 0.22 & 47.98 $\pm$ 40.19 & 1.00 $\pm$ 1.25 &   \\
375 & 80484 & 0.25066 & 3.03874 & 12.93 & 7.91 & 2.86 & 0.25 $\pm$ 0.13 & 55809.82 $\pm$ 0.71 & 72.64 $\pm$ 29.85 & 1.00 $\pm$ 0.65 & DET. \\
375 & 23911 & 0.06543 & 3.51626 & 14.55 & 6.53 & 0.74 & 0.18 $\pm$ 0.39 & 56005.30 $\pm$ 1.62 & 30.45 $\pm$ 49.30 & 0.16 $\pm$ 0.42 &   \\
375 & 90864 & 0.06159 & 3.38532 & 16.63 & 4.77 & 7.59 & 0.00 $\pm$ 2.00 & 55990.01 $\pm$ 0.85 & 20.16 $\pm$ 27.17 & 1.00 $\pm$ 2.50 &   \\
375 & 54167 & 0.07092 & 2.90685 & 13.66 & 7.29 & 0.30 & 0.70 $\pm$ 2.00 & 55998.94 $\pm$ 0.47 & 7.94 $\pm$ 135.85 & 1.00 $\pm$ 56.03 &   \\
375 & 35747 & -1.05529 & 3.12250 & 14.22 & 6.81 & 0.73 & 0.66 $\pm$ 2.00 & 57156.71 $\pm$ 135.42 & 64.06 $\pm$ 300.15 & 1.00 $\pm$ 17.69 & DET. \\
375 & 83454 & -0.84046 & 3.59040 & 16.23 & 5.11 & 2.58 & 0.00 $\pm$ 2.00 & 55998.09 $\pm$ 0.36 & 90.22 $\pm$ 89.83 & 0.17 $\pm$ 0.20 &   \\
375 & 104369 & -0.85058 & 3.55138 & 16.36 & 4.99 & 1.70 & 0.37 $\pm$ 1.28 & 56821.12 $\pm$ 1.67 & 12.28 $\pm$ 23.19 & 1.00 $\pm$ 4.37 &   \\
375 & 33561 & -0.79048 & 3.13094 & 13.35 & 7.55 & 0.63 & 0.75 $\pm$ 0.83 & 55828.49 $\pm$ 3.15 & 47.26 $\pm$ 33.76 & 1.00 $\pm$ 2.02 & DET. \\
375 & 44220 & -0.87302 & 3.83148 & 13.61 & 7.33 & 1.94 & 0.34 $\pm$ 0.38 & 56824.25 $\pm$ 1.12 & 6.34 $\pm$ 3.75 & 1.00 $\pm$ 1.71 & DET. \\
375 & 102238 & -0.35875 & 3.37052 & 14.51 & 6.56 & 1.38 & 0.00 $\pm$ 2.00 & 55999.57 $\pm$ 0.36 & 30.11 $\pm$ 48.04 & 0.32 $\pm$ 0.77 &   \\
375 & 82376 & -0.38249 & 2.86060 & 14.02 & 6.99 & 0.35 & 0.42 $\pm$ 0.37 & 56081.79 $\pm$ 4.93 & 64.92 $\pm$ 33.39 & 0.20 $\pm$ 0.26 & DET. \\
375 & 24001 & 0.05155 & 3.51611 & 13.96 & 7.03 & 1.23 & 0.28 $\pm$ 0.09 & 56866.84 $\pm$ 0.28 & 52.73 $\pm$ 6.56 & 1.00 $\pm$ 0.28 &   \\
375 & 73190 & 0.00278 & 3.42125 & 16.82 & 4.60 & 3.33 & 0.23 $\pm$ 0.22 & 56795.27 $\pm$ 3.63 & 67.86 $\pm$ 43.14 & 1.00 $\pm$ 1.22 & DET. \\
375 & 65654 & -0.42227 & 3.62051 & 14.54 & 6.54 & 0.29 & 0.52 $\pm$ 2.00 & 56877.25 $\pm$ 23.26 & 19.73 $\pm$ 207.05 & 1.00 $\pm$ 40.03 &   \\

\end{longtable*}
\footnotemark{One standard deviation errors are presented along each parameter obtained from the microlensing model fitting procedure} \\
\indent \footnotemark{Typical positional errors are 0.1 arcsec \citep{Smith17}. } \\
\indent \footnotemark{Typical photometric errors are $\sigma_{K_s} = 0.01$ mag, and $\sigma_{J, H} = 0.03$ mag \citep{saito12, contreras17, alonso18}. } \\


\end{document}